\documentclass[twocolumn]{aastex631}

\newcolumntype{Y}{>{\centering\arraybackslash}p{4.5em}}
\newcommand{\vout}{\ensuremath{v_\mathrm{out}}\xspace}
\newcommand{\nh}{\ensuremath{N_\mathrm{H}}\xspace}
\newcommand{\xmm}{\textit{XMM-Newton}\xspace}
\newcommand{\iras}{IRAS~13224$-$3809\xspace}
\newcommand{\phabs}{{\tt phabs}\xspace}
\newcommand{\pow}{{\tt powerlaw}\xspace}

\newcommand{\zxipcf}{{\tt zxipcf}\xspace}

\newcommand{\redden}{{\tt redden} \xspace}
\newcommand{\agnslim}{{\tt agnslim} \xspace}
\newcommand{\kabs}{{\tt kabs} \xspace}
\newcommand{\diskbb}{{\tt diskbb} \xspace}
\DeclareRobustCommand{\ion}[2]{\textup{#1\,\textsc{\lowercase{#2}}}}

\usepackage{threeparttable}
\usepackage{comment}
\usepackage{xspace}
\usepackage{hyperref}
\usepackage{url}
\usepackage{rotating}
\usepackage{multirow}
\usepackage{chngcntr}
\usepackage{tabularx} 
\usepackage[whole]{bxcjkjatype} 

\received{}
\revised{\today}
\accepted{}
\submitjournal{ApJ}

\shorttitle{Radiatively-driven clumpy absorbers}
\shortauthors{Midooka et al.}
\graphicspath{{./}{fig/}}

\begin{document}

\title{Radiatively-driven  clumpy X-ray absorbers in the NLS1 galaxy IRAS~13224$-$3809}

\correspondingauthor{Takuya Midooka, Misaki Mizumoto}
\email{midooka@ac.jaxa.jp, mizumoto-m@fukuoka-edu.ac.jp}

\author[0000-0002-6874-591X]{Takuya Midooka}
\affiliation{Institute of Space and Astronautical Science, Japan Aerospace Exploration Agency,\\
3-1-1 Yoshinodai, Chuo-ku, Sagamihara, Kanagawa 252-5210, Japan}
\affiliation{Department of Astronomy, Graduate School of Science, The University of Tokyo \\
7-3-1 Hongo, Bunkyo-ku, Tokyo 113-8654 Japan}

\author[0000-0003-2161-0361]{Misaki Mizumoto}
\affil{Science education research unit, University of Teacher Education Fukuoka,\\ 
1-1 Akama-bunkyo-machi, Munakata, Fukuoka 811-4192, Japan}

\author[0000-0002-5352-7178]{Ken Ebisawa}
\affiliation{Institute of Space and Astronautical Science, Japan Aerospace Exploration Agency,\\
3-1-1 Yoshinodai, Chuo-ku, Sagamihara, Kanagawa 252-5210, Japan}
\affiliation{Department of Astronomy, Graduate School of Science, The University of Tokyo \\
7-3-1 Hongo, Bunkyo-ku, Tokyo 113-8654 Japan}

\begin{abstract}
Recent radiation-magnetohydrodynamic simulations of active galactic nuclei predict the presence of the disk winds, which may get unstable and turn into fragmented  clumps far from the central black hole.
These inner winds and the outer clumps may be  observed as the ultrafast outflows (UFOs) and the  partial absorbers, respectively. However, it is challenging to observationally constrain their origins because of the complicated spectral features and variations.
To resolve such  degeneracies of the clumpy absorbers and other  components, we developed a novel   ``spectral-ratio model fitting'' technique　that estimates the  variable absorbing parameters from the ratios of the partially absorbed spectra to the non-absorbed one, canceling the complex non-variable spectral features.
We applied this method to the narrow-line Seyfert 1 galaxy \iras observed by \xmm in 2016 for $\sim$1.5~Ms.
As a result, we found that the soft spectral variation is mostly caused by changes in the partial covering fraction of the mildly-ionized clumpy absorbers, whose outflow velocities are similar to those of the UFO ($\sim$0.2--0.3~$c$). 
Furthermore, the velocities of the clumpy absorbers and UFOs increase similarly with the X-ray fluxes, consistent with the change in the UV-dominant continuum flux.
We also discovered a striking correlation between the clump covering fraction and the equivalent width of the UFO absorption lines, which indicates that increasing the outflow in the line-of-sight lead to more prominent UFOs and more partial absorption.
These findings strongly suggest that the clumpy absorbers and the UFO share  the same origin, driven by the same UV-dominant continuum radiation.
\end{abstract}

\keywords{Active galactic nuclei (16) --- Seyfert galaxies (1447) --- X-ray astronomy(1810)}

\section{Introduction}\label{sec:1} 
Active galactic nuclei (AGNs) are powered by accretion onto supermassive black holes (SMBHs).
Recent observational and theoretical studies suggest that most AGNs have multiple ionized absorbers (see review by e.g., \citealp{Laha21}). Broadly speaking, three types of absorbers have been proposed.
The first one is the {\it warm absorbers} (WAs), which are detected as absorption lines and edges in the soft X-ray band of $\sim$65\% of the nearby AGNs, especially in the radio-quiet Seyfert 1 galaxies (e.g., \citealp{McKernan07}, \citealp{Laha14}).
These absorbers are known to be mildly ionized and blue-shifted with a velocity of $\lesssim2000~ {\rm km/s}$.

The second one is the {\it partial covering absorbers}, which partially absorb the X-ray continuum in the line of sight. This is also called the ``obscurer'', which causes simultaneous soft X-ray and UV absorption troughs (e.g., NGC~5548; \citealp{Kaastra14}).

The third one is the {\it ultrafast outflow} (UFO). Seyfert galaxies often have blue-shifted, highly ionized iron K-absorption lines in their 30--40\% of the X-ray spectra \citep{Tombesi10a, Gofford13, Igo20, Matzeu23}. 
The UFOs are considered to be the outflowing winds from the accretion disk at very high velocities of 0.1--0.3~$c$, where $c$ is the speed of light. Since UFOs are likely to have larger solid angles than the relativistic jets (e.g., \citealp{Nardini15}, \citealp{Hagino15}), UFO contribution to the co-evolution of the SMBHs and the host galaxies may be comparable to or even exceed that of the jets (e.g., \citealp{King10}, \citealp{King15}).

\iras is a narrow-line Seyfert 1 (NLS1) galaxy with a SMBH of about $10^{6-7}$~M$_{\odot}$ at $z=$0.0658 (e.g., \citealp{Emmanoulopoulos14, Chiang15, Alston19}), accreting close to or above the Eddington limit \citep{Alston19}.
In this object, \cite{Parker17} discovered UFO absorption lines in the 1.5~Ms deep observation by \xmm in 2016. They claimed that the equivalent width of the UFO absorption lines and the X-ray luminosity are anti-correlated, while the line-of-sight velocity of the UFO is correlated with the X-ray luminosity. 

Some previous studies including \cite{Parker17} explained the X-ray spectra of \iras in terms of the relativistic disk reflection model (e.g., \citealp{Fabian89}), which assumes such extreme physical conditions that the central black hole spin is almost maximum and the most incident X-ray radiation from a tiny source is reflected at the innermost region of the disk (e.g., \citealp{Fabian09}). 
This model also requires an iron overabundance by a factor of 3--20 to account for energy spectra of various objects (e.g., \citealp{Fabian13}, \citealp{Chiang15}).
In a recent study, \citet{Jiang22a} reported that the iron abundance $Z_{\rm Fe}$ is $3.2\pm0.5$ when assuming a high-density disk reflection of 10$^{20}$~cm$^{-3}$.
In the X-ray spectra of AGN, the relativistic reflection model and the partial absorption model are often indistinguishable in shape (e.g., \citealp{Parker22}). In fact, \cite{Yamasaki16} successfully explained the spectral variability of \iras (\xmm data in 2001--2011 before detection of the UFO absorption) using the partial absorption model without such extreme conditions.

Two-dimensional radiation-magnetohydrodynamic simulations of supercritical accretion flows show that the radiation pressure generates disk winds, which become clumpy at a few hundred Schwarzschild radius (R$_{\rm s}$) \citep{Takeuchi13}.
They argued that the clump formation is probably because the Rayleigh-Taylor instability works efficiently where the radiation pressure is dominant over the gravitational potential. In addition to the Rayleigh-Taylor instability, contribution of the radiation hydrodynamic instability to the clump formation is also suggested in two- or three-dimensional radiation hydrodynamic simulations (e.g., \citealp{Takeuchi14}, \citealp{Kobayashi18}).
Another mechanism has been proposed for the disk wind to get clumpy:
\cite{Dannen20} found through a parsec-scale wind simulation that dynamical thermal instability occurs in some zones, causing fragmentation of the outflow beyond the acceleration radius of the wind. 

\cite{Mizumoto19} proposed the ``hot inner and clumpy outer wind model'', in which the inner wind and outer clumps may actually be observed as the UFOs and the clumpy absorbers, respectively.
However, in the standard X-ray spectral analysis, parameters of the clumpy absorbers and other components such as WAs are often degenerate and difficult to be disentangled.

Recently, we developed a new data analysis technique called ``spectral-ratio model fitting'' to disentangle the spectral parameter degeneracy (\citealp{Midooka22a}, hereafter Paper~1). 
In this method,  by taking the spectral ratios of the intensity-sliced spectra, spectral variations only due to change of the partial absorbers are manifested,  canceling out the invariable continuum and absorption features.

In this paper, we aim to constrain the outflowing velocity of the clumpy absorbers of \iras using the spectral ratio fitting method, and search for a plausible origin of the outflow and clumps.

\begin{figure}
\centerline{\includegraphics[width=1.0\columnwidth]{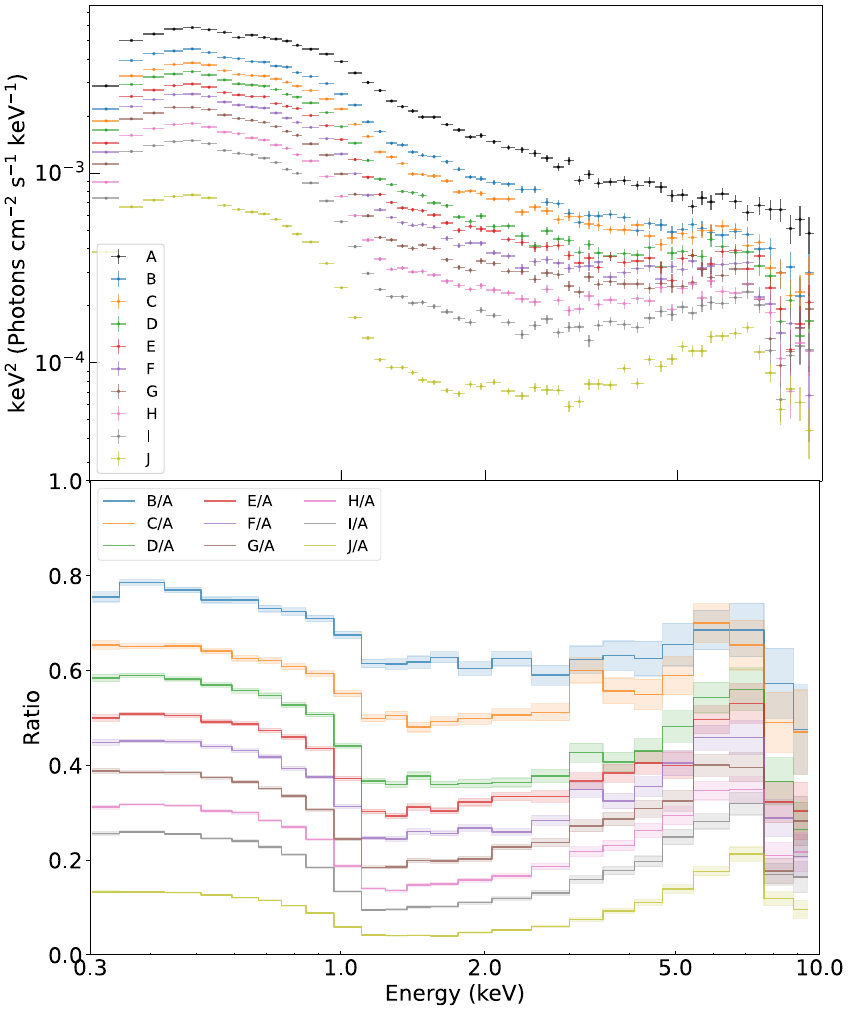}}
	    \caption{Top panel shows the 10 intensity-sliced spectra obtained by \xmm pn in 0.3--10.0~keV. The $\nu F_\nu$ plot unfolded by the power-law with the index of 2.0 is shown. The bottom panel shows the nine spectral ratios to the brightest spectrum A (black in the top panel). Note that both panels show  adequately binned spectra to enhance visibility.}	   \label{fig:sliced}
\end{figure}

\section{Observations and Methods}\label{sec:2}
\subsection{Observations and Data reduction}
We used the 1.5~Ms data obtained by \xmm EPIC-pn (\citealp{Jansen01}, \citealp{Struder01}) and OM \citep{Mason01} in the summer of 2016, including the unscheduled observations (see Table~1 in Paper~1).
As our study requires  a consecutive dataset taken within a short period of time, we did not incorporate any archival data before 2016.
In addition to the pn, of which data processing was described in Paper~1, the OM data taken with the UVW1 filter is reduced with the standard SAS routine {\tt omichain}, and the spectrum is extracted using {\tt om2pha}.

Since the UFO properties of \iras are known to be luminosity-dependent (e.g., \citealp{Parker17}, \citealp{Pinto18}, and \citealp{Chartas18}), we created the intensity-sliced spectra.
First, all the 0.3–10.0~keV events were binned into 1~ks intervals, as in previous studies \citep{Parker17, Pinto18}. We grouped the data into 10 intensity levels based on the flux in the 0.3--10~keV range and generated 10 intensity-sliced spectra
(see Table~2 in Paper~1), which are labeled from A (the brightest) to J (the dimmest).
The sliced spectral bins are grouped with 8 and 16  bins in 0.3--2.0~keV and 2.0--10.0~keV, respectively.
These intensity-sliced spectra are shown in the upper panel of Figure~\ref{fig:sliced}.
Next, the observed spectral ratios are created by taking the ratio of each intensity-sliced spectrum (B--J) to the brightest spectrum A (lower panel of Figure~\ref{fig:sliced}).

\subsection{Method of the spectral-ratio model fitting} 
We briefly describe the spectral-ratio model fitting method,
details of which were explained in Paper~1.
The spectral continuum of \iras consists of the power-law (PL) and the soft-excess component, and the continuum is absorbed by several intervening gases such as WA, UFO, and clumpy absorbers.
WAs do not change their geometry or ionization structure significantly in a timescale of a few weeks, while the partial covering fraction of the clumpy absorbers, whose variation influences the spectral shape below $\sim$5~keV, significantly changes in this timescale (e.g., \citealp{DiGesu15}, \citealp{Midooka22}).
Therefore, taking the spectral ratio below 5~keV enables us to focus on the variability of the clumpy absorbers, canceling out the less time-variable spectral components. 
It should be noted that a similar spectral-ratio technique has been successfully adopted to study  variation of the high energy cutoff in the  AGN spectra \citep{Zhang18}.

Model spectral ratios in 0.3--5.0~keV are calculated using XSPEC (version 12.12.0).
We consider a simple model spectrum in which the continuum is covered by an ionized partial absorber (\zxipcf; \citealp{Reeves08}).
This absorber model has three parameters; ionization parameter $\xi$, hydrogen column density \nh, and partial covering fraction (CF). 
A table model of the spectral ratio is created by taking the ratio of the absorbed model spectra to the non-absorbed one (CF$=$0) with the same continuum parameters (see Figure~2 in Paper~1). 

The three parameters of the clumpy absorber determine the shape of the spectral ratio, including the characteristic dip structures in 0.8--1.0~keV (Figure~2 in Paper~1).
Fitting the observed and model spectral ratios makes it possible to constrain the partial absorber parameters without being affected by
complicated WA spectral features.

\begin{figure}
\centerline{\includegraphics[width=0.9\columnwidth]{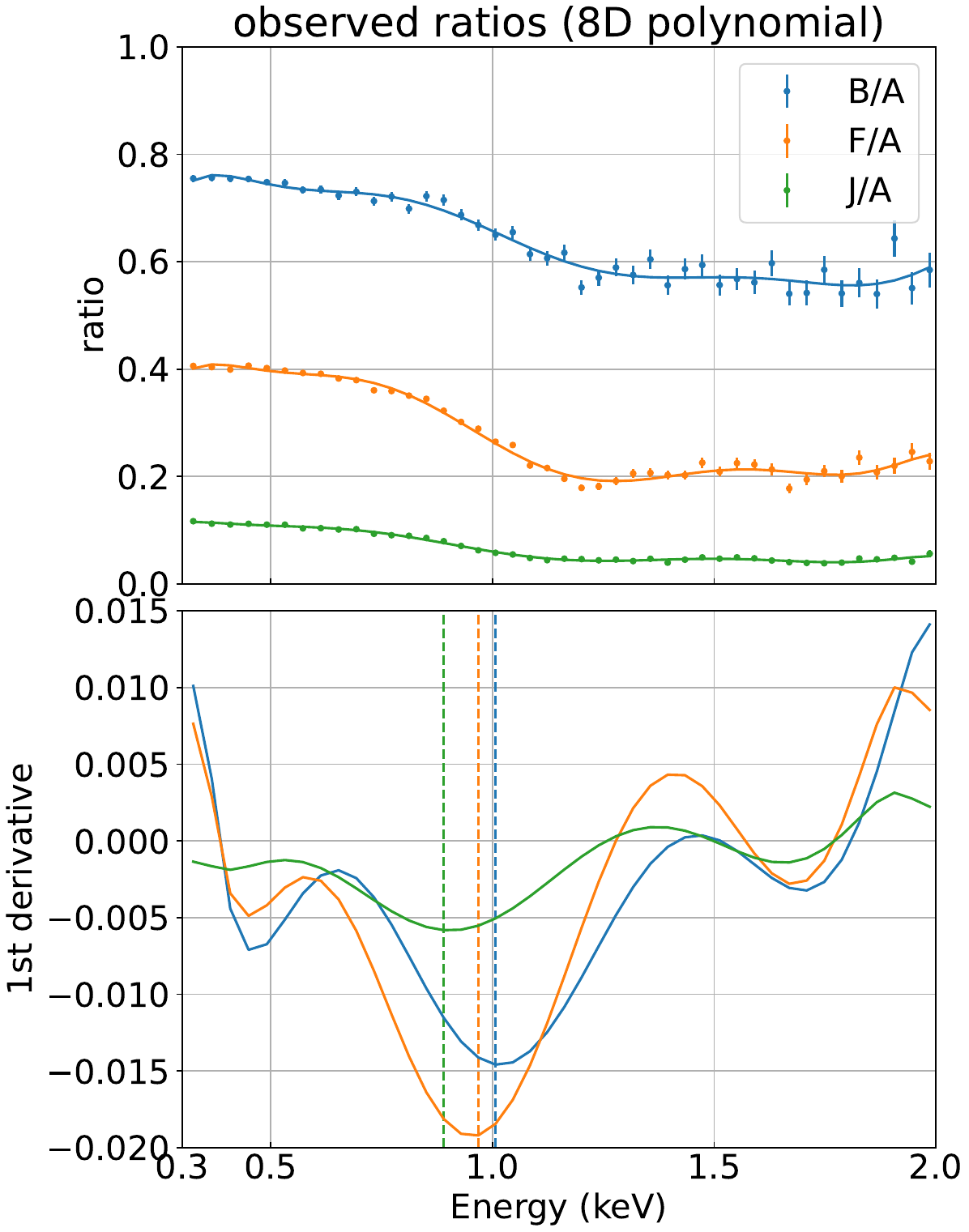}}
	    \caption{In the upper panel, scattered points show the observed spectral ratios B/A, F/A, and J/A. Solid lines are the best-fit 8th-order polynomial equations. The lower panel shows their 1st derivatives, in which the dotted vertical lines represent the energies where the 1st derivatives get minimum. These energies, characterizing the dip feature, increases as the X-ray flux gets higher.} 
	    \label{fig:1stderi}
\end{figure}

\begin{figure*}
\centerline{\includegraphics[width=1.8\columnwidth]{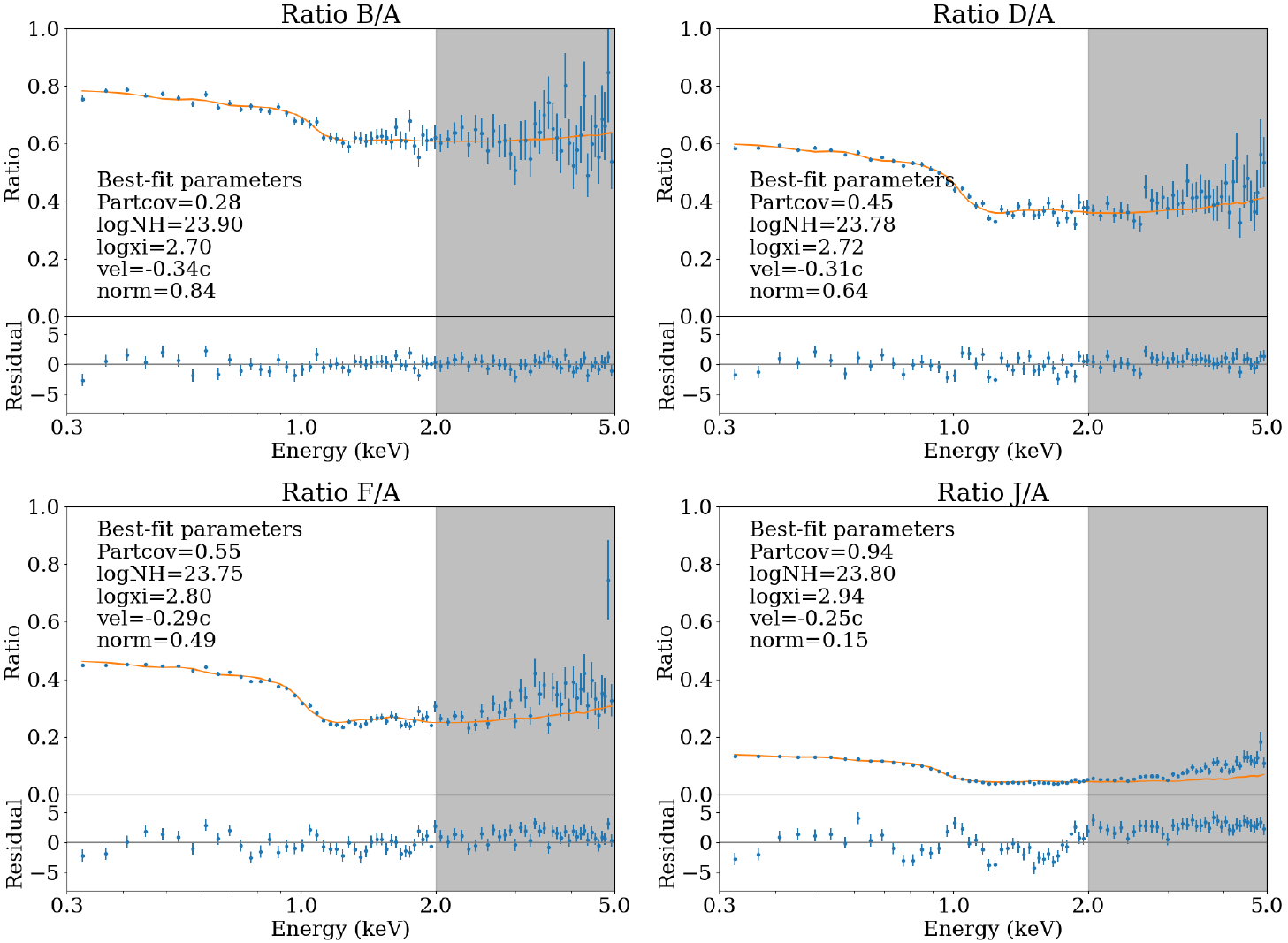}}
	    \caption{Examples of the spectral-ratio model fitting with the least chi-square method (B/A, D/A, F/A, and J/A). Fitting for the dimmer ones shows more significant residuals above $\sim$2~keV, suggesting changes in the PL index, which is not taken into account here. The absorber parameters are constrained by the fitting below 2~keV, so the energy bands above 2~keV are shaded in gray.}
	    \label{fig:BDFJ_ratiofit}
\end{figure*}

\section{Data Analysis and Results}\label{sec:3}
\subsection{Model-independent study of the spectral ratios} \label{sec:3.1}
One of the purposes of the present study is to constrain the clump outflow velocities. 
In this section, we show that the flux dependency of the clump velocity can be constrained model-independently from the spectral ratios. 

We made 9 spectral ratios from the 10 observed intensity-sliced spectra.
The upper panel in Figure~\ref{fig:1stderi} shows the representative spectral ratios of B/A, F/A, and J/A, each of which has a dip- or cliff-like structure at around 1~keV. 
We define the ``dip characteristic energy'' where the spectral ratio changes from convex-upward to convex-downward.
To do so, first, the observed spectral ratios are fitted with 8th-order polynomial equations (solid lines in the upper-panel in Figure~\ref{fig:1stderi}). 
Then, the first derivatives of the 8th-order equation models are calculated (lower panel in Figure~\ref{fig:1stderi}).
The energies that give the minimums of the first derivative correspond to the dip characteristic energies defined above (dotted vertical lines in the lower panel of Figure~\ref{fig:1stderi}).
We find that the dip characteristic energy is {\em blue-shifted} when X-rays get brighter.
Note that this result is obtained without assuming any spectral models.

We note that the relativistic reflection model (e.g., \citealp{Parker17}, \citealp{Pinto18}, \citealp{Chartas18}, and \citealp{Jiang22a}) can  also reproduce these spectral ratios, because the individual spectra are fitted.
However, changes in the  continuum and/or disk reflection normalizations are insufficient
to account for the observed blue-shift of the dip structure, and the disk geometry and other parameters need to change concordantly to realize such spectral changes.  We do not discuss this possibility any further in this paper.

\begin{figure}
\centerline{\includegraphics[width=0.95\columnwidth]{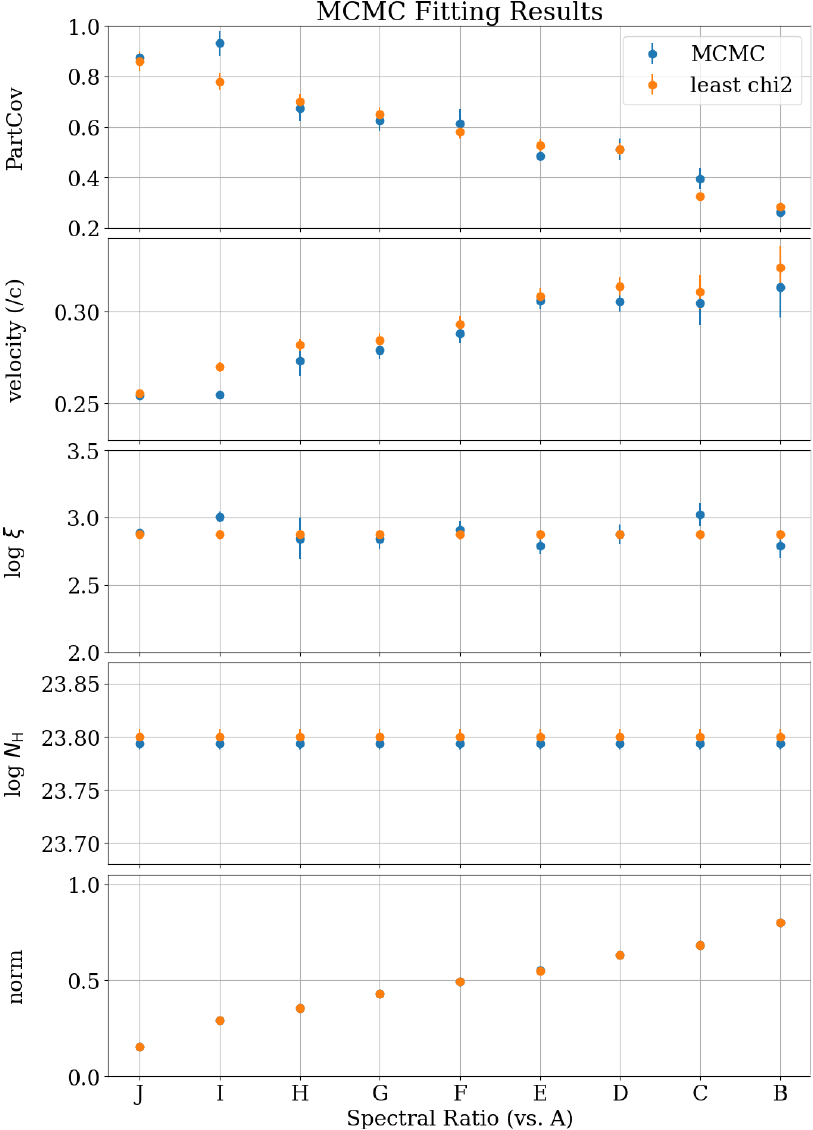}}
	    \caption{Best-fit parameters determined by the simultaneous least chi-square fitting (in orange) and MCMC estimation (in blue) to the nine intensity-sliced spectral ratios. \nh and $\xi$ are tied in the least chi-square fitting, and only \nh is tied in the MCMC for all the spectral ratios. 
	    These parameters are indicated in Table~\ref{tab:ratio_par}.}
	    \label{fig:bestparams}
\end{figure}

\begin{table*}
    \centering
    \caption{Best-fit parameters obtained from spectral ratio model fitting with the least chi-squared (upper) and the MCMC estimation (lower). Note that \nh and $\xi$ are tied in the least chi-square fitting, and only \nh is tied in the MCMC for all the spectral ratios.} 
    \label{tab:ratio_par}
\begin{tabular}{llllllllll}
\textbf{Chi-sq} & \multicolumn{1}{r}{J} & \multicolumn{1}{r}{I} & \multicolumn{1}{r}{H} & \multicolumn{1}{r}{G} & \multicolumn{1}{r}{F} & \multicolumn{1}{r}{E} & \multicolumn{1}{r}{D} & \multicolumn{1}{r}{C} & \multicolumn{1}{r}{B} \\ \hline
CF & 0.86 & 0.78 & 0.70 & 0.65 & 0.58 & 0.53 & 0.51 & 0.32 & 0.28 \\
norm & 0.15 & 0.29 & 0.35 & 0.43 & 0.50 & 0.55 & 0.63 & 0.68 & 0.80 \\
log \nh (tied) & 23.8 & 23.8 & 23.8 & 23.8 & 23.8 & 23.8 & 23.8 & 23.8 & 23.8 \\
log $\xi$ (tied) & 2.87 & 2.87 & 2.87 & 2.87 & 2.87 & 2.87 & 2.87 & 2.87 & 2.87 \\
velocity (/c) & 0.26 & 0.27 & 0.28 & 0.28 & 0.29 & 0.31 & 0.31 & 0.31 & 0.32  \\ \hline
\end{tabular}
\end{table*}

\begin{table*}
    \begin{center}
\begin{tabular}{llllllllll}
\textbf{MCMC} & \multicolumn{1}{r}{J} & \multicolumn{1}{r}{I} & \multicolumn{1}{r}{H} & \multicolumn{1}{r}{G} & \multicolumn{1}{r}{F} & \multicolumn{1}{r}{E} & \multicolumn{1}{r}{D} & \multicolumn{1}{r}{C} & \multicolumn{1}{r}{B} \\ \hline
CF & 0.88 & 0.93 & 0.67 & 0.62 & 0.61 & 0.48 & 0.51 & 0.40 & 0.26 \\
norm & 0.15 & 0.29 & 0.36 & 0.43 & 0.50 & 0.55 & 0.63 & 0.68 & 0.80 \\
log \nh (tied) & 23.8 & 23.8 & 23.8 & 23.8 & 23.8 & 23.8 & 23.8 & 23.8 & 23.8 \\
log $\xi$ & 2.89 & 3.01 & 2.84 & 2.84 & 2.91 & 2.79 & 2.88 & 3.02 & 2.79 \\
velocity (/c) & 0.25 & 0.25 & 0.27 & 0.28 & 0.29 & 0.31 & 0.31 & 0.30 & 0.31  \\ \hline
\end{tabular}
    \end{center}
\end{table*}

\subsection{Spectral-ratio model fitting in 0.3--5~keV} \label{sec:3.2}
Next, we are going to fit the observed spectral ratios with physical models to constrain absorbers' parameters. Having noticed the blue-shifted absorption features around 1~keV, we introduce the clumpy absorber velocity \vout in addition to the three clump parameters (CF, \nh, and $\xi$). 
Also, the continuum normalization ratio is a free parameter.

Spectral-ratio table model was created with these four free parameters (CF, $\xi$, \nh, and \vout; see Table~3 in Paper~1), where models are linearly interpolated between the adjacent grid points using the XSPEC built-in function. 
First, the least-square fitting was performed for the nine observed spectral ratios allowing all the five parameters free (the four parameters above and the normalization).
We found that the \nh and $\xi$ were not significantly variable among the nine ratios ($2.7 < {\rm log}~\xi < 3.0$, $23.7 <$ log~\nh $< 23.9$). 
Thus we tied these two parameters for all the nine spectral ratios and performed simultaneous fitting. 

Table \ref{tab:ratio_par} shows all the fitting results,
and Figure~\ref{fig:BDFJ_ratiofit} shows examples of the fitting B/A, D/A, F/A, and J/A.
The fitting was successful in all the spectral ratios below $\sim$2~keV, where the absorbers' parameters are constrained.
Note that the model does not fit well above $\sim$3~keV, especially in dimmer spectral ratios, suggesting a change in the photon index, that is confirmed later in the spectral fitting (section \ref{sec:fitting}).
The best-fit parameters are illustrated in Figure~\ref{fig:bestparams}.
We found that the partial covering fraction is smaller and the norm is higher as the X-ray flux gets higher.
The outflow velocity of the partial absorber was 0.2--0.3~$c$, unexpectedly fast, even comparable to that of the UFOs reported in previous studies \citep{Parker17, Pinto18, Chartas18}. Furthermore, the velocity was found to be higher when the X-ray flux is higher.

We aim to confirm that the velocity we have obtained is determined independently, not being degenerate with the ionization parameter $\xi$, since both parameters similarly affect absorption line/edge energies.
We perform a Markov Chain Monte Carlo (MCMC) calculation using the ``emcee'' package \citep{Foreman-Mackey13} to determine the posterior distribution of the best-fit spectral parameters.
The initial parameters in the MCMC chains are set to be close to the best-fit parameters obtained by the chi-squared fitting.
We set a uniform prior distribution over the parameter ranges (Table~3 in Paper~1).
After the initial 5000 steps are discarded to exclude the burn-in phase, further 10000 steps are explored by 1000 separate chains (``walkers''). 

We succeeded in the MCMC parameter estimation for all the spectral ratios.
The best-fit model ratios are shown in Appendix (Figure~\ref{fig:all_ratiofitting}).
Blue dots in Figure~\ref{fig:bestparams} show the mean values obtained by the MCMC calculation.
We confirmed that the best-fit parameters and their flux-dependency are similar to those by the chi-squared fitting (orange dots in Figure~\ref{fig:bestparams}). Note that $\xi$ is not tied among the different ratios unlike the least chi-square fitting, because the primary aim of this analysis is to ensure whether there is no correlation between \vout and $\xi$.

\begin{figure}
\centerline{\includegraphics[width=1.0\columnwidth]{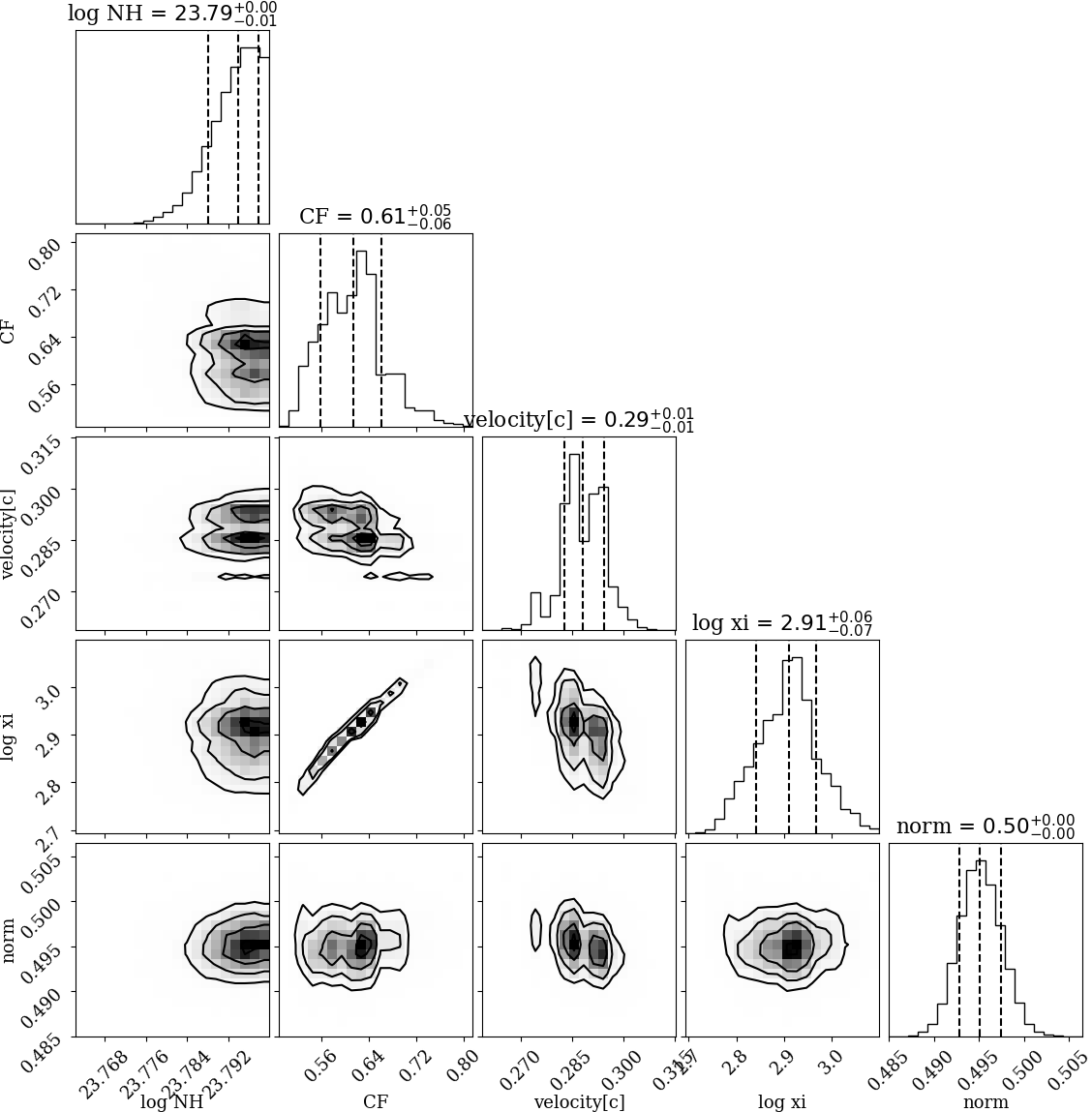}}
	    \caption{A corner plot showing the result of the MCMC parameter estimation for the typical spectral ratio F/A. The velocity is not correlated with other parameters, whereas a correlation is found between CF and log $\xi$.}
	    \label{fig:contour}
\end{figure}

Figure~\ref{fig:contour} shows a corner plot of the two-dimensional correlations between the posterior distributions of the estimated parameter pairs in the spectral ratio F/A.
The other corner plots are shown in Appendix, Figure~\ref{fig:all_contours}.
We found that the velocity is not correlated with other parameters and is determined independently.
In contrast, CF and $\xi$ are correlated in most spectral ratios. Still, whereas the CF shows a clear anti-correlation with the X-ray flux, the $\xi$ variation is tiny and will not affect the intensity variations (Figure \ref{fig:bestparams}). 

\subsection{Spectral fitting in the 0.3--10~keV band}\label{sec:fitting} 
We performed spectral fitting in 0.3--10~keV to constrain the UFO velocity with the partial absorption model based on the ratio model fitting.
The spectral data are binned to have at least 20 counts per new energy bin.
The XSPEC spectral model adopted is represented as
\begin{equation} 
    \phabs*\zxipcf*(\pow + \diskbb)*{\tt const}*\kabs,
\end{equation}
where each spectral component is explained below.

The column density of \phabs for the interstellar absorption was fixed at the Galactic value \nh $= 5\times 10^{20}$ cm$^{-2}$ \citep{Bekhti16}, where we set the cosmic abundances to the {\tt wilms} values \citep{Wilms00a}.
In addition to the model for the ratio fitting, the \kabs model (\citealp{Ueda04}, updated by \citealp{Tomaru20}) is incorporated to explain the \ion{Fe}{XXV} and/or \ion{Fe}{XXVI} UFO absorption features.
The \kabs model calculates the Voigt profile absorption lines (K$\alpha $ and K$\beta$) for \ion{Fe}{XXV} and \ion{Fe}{XXVI} for the given column density of the ion $N_{\rm atom}$ and the turbulent velocity $v_{\rm turb}$. The redshift $z$ due to the line-of-sight velocity is also a free parameter.
Although \cite{Pinto18} reported the presence of \ion{O}{VIII} UFO absorption line in the RGS spectra, we do not include the absorption line in the spectral model for simplicity.
Also, disk  reflection components are not required  in our spectral model.

Initial parameters of the spectral fitting are set to the values constrained by the ratio model fitting.
Both the continuum parameters and the partial absorption parameters were free to vary for the fitting.
It is difficult to place limits on the ionization parameter and the turbulent velocity $v_{\rm turb}$ simultaneously from the absorption line profiles above 8~keV. Therefore, $v_{\rm turb}$ was fixed at 10000~km/s, and the fitting was performed assuming that the UFO absorption lines are produced either by He-like Fe or H-like Fe.

\begin{figure}
\centerline{\includegraphics[width=1.1\columnwidth]{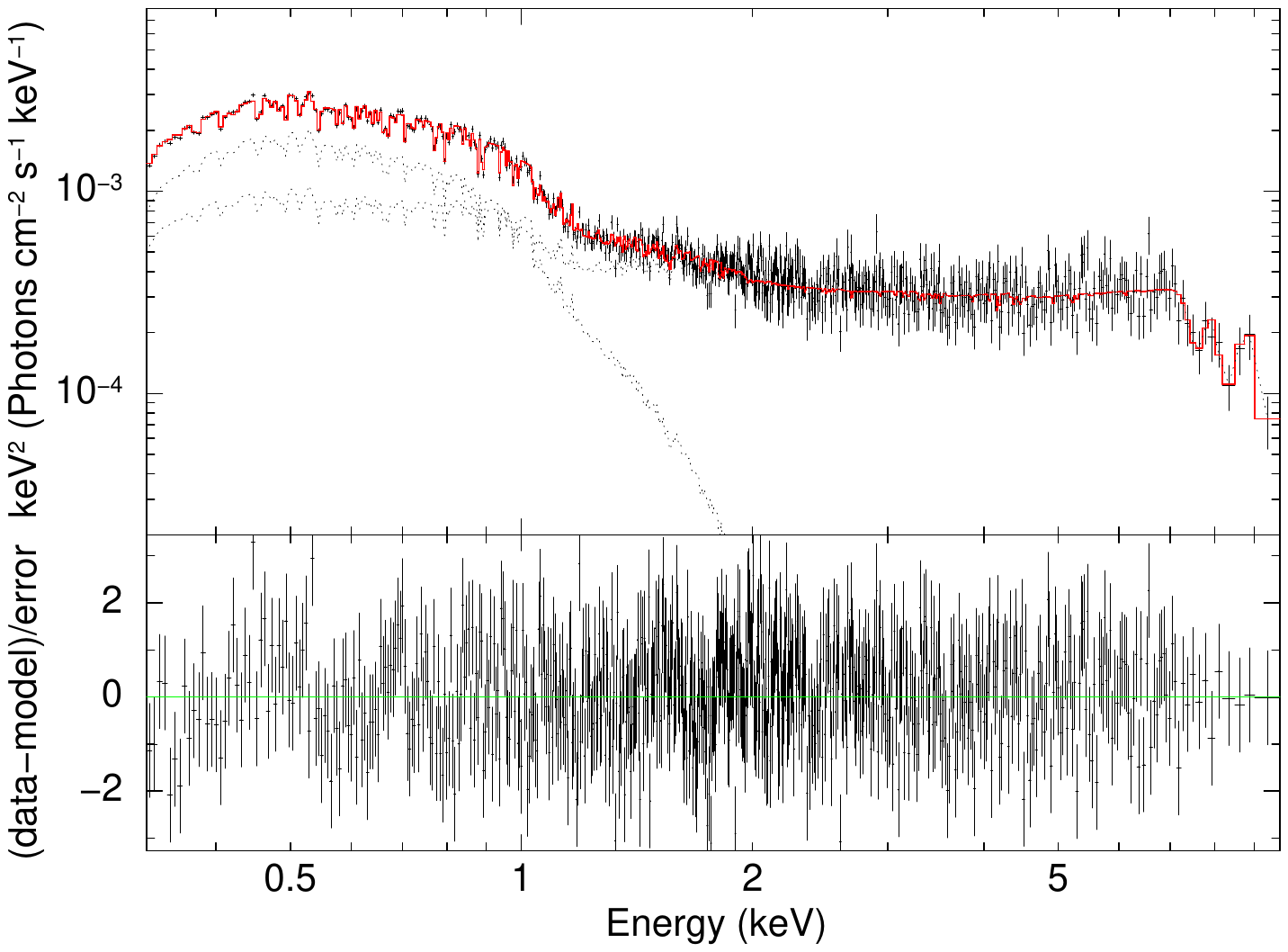}} 
	    \caption{Spectral fitting result for the typical spectrum F in 0.3--10~keV. The dotted lines in black show \pow and {\tt diskbb}, respectively. The red line represents the best-fit model, including three {\tt kabs} lines at 7--10~keV. The lower panel shows the residuals of the model fitting. The other fitting results are shown in Appendix (Figure~\ref{fig:all_specfit}).} 
	    \label{fig:F_specfit}
\end{figure}

In the spectral fitting, we include K$\alpha$ and K$\beta$ lines as a pair. 
Three K$\alpha$ lines (accompanied by three K$\beta$ lines) at maximum are necessary to fit the UFO feature depending on the spectrum.
This indicates that more than one velocity component of the UFO was observed.
The fitting result of a typical intensity-sliced spectrum F is shown in Figure~\ref{fig:F_specfit}, while the others are shown in Figure~\ref{fig:all_ratiofitting}.
Best-fit parameters are given in Table~\ref{tab:best_specfit}.

Note that our previous study based on a similar partial covering model required a strong absorption edge around 1~keV to explain the residual feature \citep{Yamasaki16}. Now, the fit is successful without such an additional edge, which suggests that the residual feature was caused by the  energy-shift of the  {\em outflowing}\/ clumpy absorbers.

\begin{table*} 
    \centering
    \caption{Best-fit parameters determined by the spectral fitting in 0.3--10.0~keV. 
    The CF of \zxipcf and {\tt const} are fixed to the  values determined by the chi-squared ratio fitting (upper panel of Table~\ref{tab:ratio_par}).
    Fitting was made  assuming that the UFO absorption lines are produced either by He-like Fe or H-like Fe.
    The parameters except \kabs are determined assuming  the He-like Fe. 
    Fitting results are shown in Figure~\ref{fig:all_specfit}.  Hereafter the error ranges correspond to the 90\% confidence level.}
	\label{tab:best_specfit}
    \begin{tabularx}{150mm}{ccccccc}\hline
Components & Parameters & A & B & C & D & E \\\hline
\zxipcf & \nh (10$^{22}$cm$^{-2}$) & --- & $36^{+13}_{-7}$ & $46^{+9}_{-6}$ & $51^{+4}_{-5}$ & $56^{+5}_{-5}$ \\
& log $\xi$ & --- & $2.72^{+0.04}_{-0.04}$ & $2.73^{+0.03}_{-0.05}$ & $2.73^{+0.05}_{-0.11}$ & $2.72^{+0.03}_{-0.03}$ \\
& z & --- & $-0.33^{+0.01}_{-0.02}$ & $-0.33^{+0.01}_{-0.01}$ & $-0.32^{+0.01}_{-0.01}$ & $-0.32^{+0.01}_{-0.01}$ \\
\pow & $\Gamma$ & $2.67^{+0.03}_{-0.03}$ & $2.71^{+0.03}_{-0.03}$ & $2.60^{+0.03}_{-0.03}$ & $2.60^{+0.03}_{-0.04}$ & $2.54^{+0.03}_{-0.03}$ \\
& norm (10$^{-3}$) & $2.2^{+0.1}_{-0.1}$ & $2.3^{+0.1}_{-0.1}$ & $2.2^{+0.1}_{-0.1}$ & $2.3^{+0.1}_{-0.2}$ & $2.4^{+0.1}_{-0.1}$ \\ 
\diskbb & $kT_{\rm in}$ (eV) & $182^{+1}_{-1}$ & $178^{+2}_{-2}$ & $177^{+2}_{-2}$ & $179^{+2}_{-3}$ & $177^{+2}_{-2}$ \\
& norm & $731^{+34}_{-33}$ & $757^{+45}_{-62}$ & $845^{+46}_{-55}$ & $781^{+89}_{-57}$ & $831^{+72}_{-59}$ \\ \hline
\multicolumn{7}{c}{{\it Assuming  \ion{Fe}{XXV}}} \\
\kabs & $N_{\rm \ion{Fe}{XXV}}$ (10$^{18}$cm$^{-2}$) & --- & $2^{+2}_{-1}$ & $4^{+2}_{-1}$ & $4^{+2}_{-1}$ & $5^{+2}_{-2}$ \\
& z & --- & $-0.26^{+0.04}_{-0.10}$ & $-0.29^{+0.03}_{-0.02}$ & $-0.24^{+0.01}_{-0.02}$ & $-0.25^{+0.02}_{-0.02}$ \\
\kabs & $N_{\rm \ion{Fe}{XXV}}$ (10$^{18}$cm$^{-2}$) & --- & --- & --- & $8^{+6}_{-3}$ & $6^{+5}_{-2}$ \\
& z & --- & --- & --- & $-0.33^{+0.01}_{-0.01}$ & $-0.32^{+0.02}_{-0.02}$ \\
\kabs & $N_{\rm \ion{Fe}{XXV}}$ (10$^{18}$cm$^{-2}$) & --- & --- & --- & --- & --- \\
& z & --- & --- & --- & --- & --- \\\hline
\multicolumn{7}{c}{{\it Assuming  \ion{Fe}{XXVI}}} \\
\kabs & $N_{\rm \ion{Fe}{XXVI}}$ (10$^{18}$cm$^{-2}$) & --- & $4^{+4}_{-3}$ & $7^{+5}_{-3}$ & $8^{+4}_{-3}$ & $10^{+5}_{-4}$ \\
& z & --- & $-0.23^{+0.04}_{-0.11}$ & $-0.26^{+0.03}_{-0.02}$ & $-0.21^{+0.01}_{-0.02}$ & $-0.22^{+0.02}_{-0.02}$ \\
\kabs & $N_{\rm \ion{Fe}{XXVI}}$ (10$^{18}$cm$^{-2}$) & --- & --- & --- & $17^{+12}_{-7}$ & $12^{+9}_{-5}$ \\
& z & --- & --- & --- & $-0.30^{+0.01}_{-0.01}$ & $-0.29^{+0.02}_{-0.01}$  \\
\kabs & $N_{\rm \ion{Fe}{XXVI}}$ (10$^{18}$cm$^{-2}$) & --- & --- & --- & --- & --- \\
& z & --- & --- & --- & --- & --- \\\hline
 $\chi^2$/dof & & 760.5/677 & 731.3/628 & 715.4/639 & 616.7/608 & 747.9/619 \\\hline
\end{tabularx}
    \begin{tabularx}{150mm}{ccccccc}\hline
Components & Parameters & F & G & H & I & J \\\hline
\zxipcf & \nh (10$^{22}$cm$^{-2}$) & $52^{+6}_{-5}$ & $56^{+5}_{-5}$ & $54^{+4}_{-4}$ & $57^{+4}_{-4}$ & $60^{+3}_{-3}$ \\
& log $\xi$ & $2.76^{+0.04}_{-0.16}$ & $2.78^{+0.03}_{-0.03}$ & $2.81^{+0.03}_{-0.03}$ & $2.85^{+0.02}_{-0.02}$ & $2.91^{+0.02}_{-0.02}$ \\
& z & $-0.31^{+0.01}_{-0.01}$ & $-0.31^{+0.01}_{-0.01}$ & $-0.30^{+0.01}_{-0.01}$ & $-0.29^{+0.01}_{-0.01}$ & $-0.28^{+0.00}_{-0.01}$ \\ 
\pow & $\Gamma$ & $2.47^{+0.04}_{-0.04}$ & $2.40^{+0.04}_{-0.04}$ & $2.34^{+0.04}_{-0.04}$ & $2.24^{+0.05}_{-0.05}$ & $1.95^{+0.05}_{-0.05}$ \\
& norm (10$^{-3}$) & $2.2^{+0.3}_{-0.2}$ & $2.2^{+0.2}_{-0.2}$ & $2.0^{+0.2}_{-0.2}$ & $1.8^{+0.2}_{-0.2}$ & $1.2^{+0.1}_{-0.1}$ \\ 
\diskbb & $kT_{\rm in}$ (eV) & $173^{+2}_{-2}$ & $170^{+2}_{-2}$ & $168^{+2}_{-1}$ & $165^{+1}_{-1}$ & $165^{+1}_{-1}$ \\
& norm & $986^{+85}_{-99}$ & $1085^{+85}_{-89}$ & $1238^{+83}_{-83}$ & $1406^{+77}_{-78}$ & $1589^{+63}_{-62}$ \\\hline
\multicolumn{7}{c}{{\it Assuming \ion{Fe}{XXV}}} \\
\kabs & $N_{\rm \ion{Fe}{XXV}}$ (10$^{18}$cm$^{-2}$) & $3^{+1}_{-1}$ & $2^{+1}_{-1}$ & $8^{+3}_{-2}$ & $3^{+1}_{-1}$ & $7^{+2}_{-1}$ \\
& z & $-0.18^{+0.01}_{-0.02}$ & $-0.14^{+0.02}_{-0.02}$ & $-0.24^{+0.01}_{-0.01}$ & $-0.20^{+0.01}_{-0.01}$ & $-0.22^{+0.01}_{-0.01}$ \\
\kabs & $N_{\rm \ion{Fe}{XXV}}$ (10$^{18}$cm$^{-2}$) & $5^{+2}_{-1}$ & $10^{+3}_{-2}$ & $6^{+4}_{-2}$ & $7^{+3}_{-2}$ & $7^{+4}_{-2}$ \\
& z & $-0.25^{+0.01}_{-0.01}$ & $-0.23^{+0.01}_{-0.01}$ & $-0.32^{+0.01}_{-0.02}$ & $-0.26^{+0.01}_{-0.01}$ & $-0.31^{+0.01}_{-0.01}$ \\
\kabs & $N_{\rm \ion{Fe}{XXV}}$ (10$^{18}$cm$^{-2}$) & $9^{+26}_{-3}$ & $5^{+3}_{-2}$ & --- & $3^{+2}_{-1}$ & --- \\ 
& z & $-0.34^{+0.01}_{-0.04}$ & $-0.31^{+0.02}_{-0.02}$ & --- & $-0.33^{+0.02}_{-0.02}$ & --- \\ \hline
\multicolumn{7}{c}{{\it Assuming \ion{Fe}{XXVI}}} \\
\kabs & $N_{\rm \ion{Fe}{XXVI}}$ (10$^{18}$cm$^{-2}$)  & $6^{+3}_{-2}$ & $5^{+2}_{-2}$ & $16^{+7}_{-4}$ & $7^{+3}_{-3}$ & $16^{+5}_{-3}$ \\
& z & $-0.14^{+0.02}_{-0.02}$ & $-0.10^{+0.02}_{-0.02}$ & $-0.21^{+0.01}_{-0.01}$ & $-0.16^{+0.01}_{-0.01}$ & $-0.19^{+0.01}_{-0.01}$ \\
\kabs & $N_{\rm \ion{Fe}{XXVI}}$ (10$^{18}$cm$^{-2}$)  & $11^{+5}_{-3}$ & $20^{+7}_{-4}$ & $13^{+9}_{-5}$ & $15^{+7}_{-5}$ & $16^{+8}_{-4}$ \\
& z & $-0.22^{+0.01}_{-0.01}$ & $-0.20^{+0.01}_{-0.01}$ & $-0.29^{+0.01}_{-0.02}$ & $-0.22^{+0.01}_{-0.01}$ & $-0.29^{+0.01}_{-0.01}$ \\
\kabs & $N_{\rm \ion{Fe}{XXVI}}$ (10$^{18}$cm$^{-2}$)  & $18^{+45}_{-6}$ & $10^{+7}_{-4}$ & --- & $6^{+4}_{-3}$ & --- \\
& z & $-0.31^{+0.01}_{-0.03}$ & $-0.28^{+0.02}_{-0.02}$ & --- & $-0.30^{+0.02}_{-0.02}$ & --- \\\hline
 $\chi^2$/dof & & 676.4/625 & 712.9/624 & 766.9/629 & 665.8/612 & 779.4/642 \\\hline
    \end{tabularx}
\end{table*}

\begin{figure}
\centerline{\includegraphics[width=1.0\columnwidth]{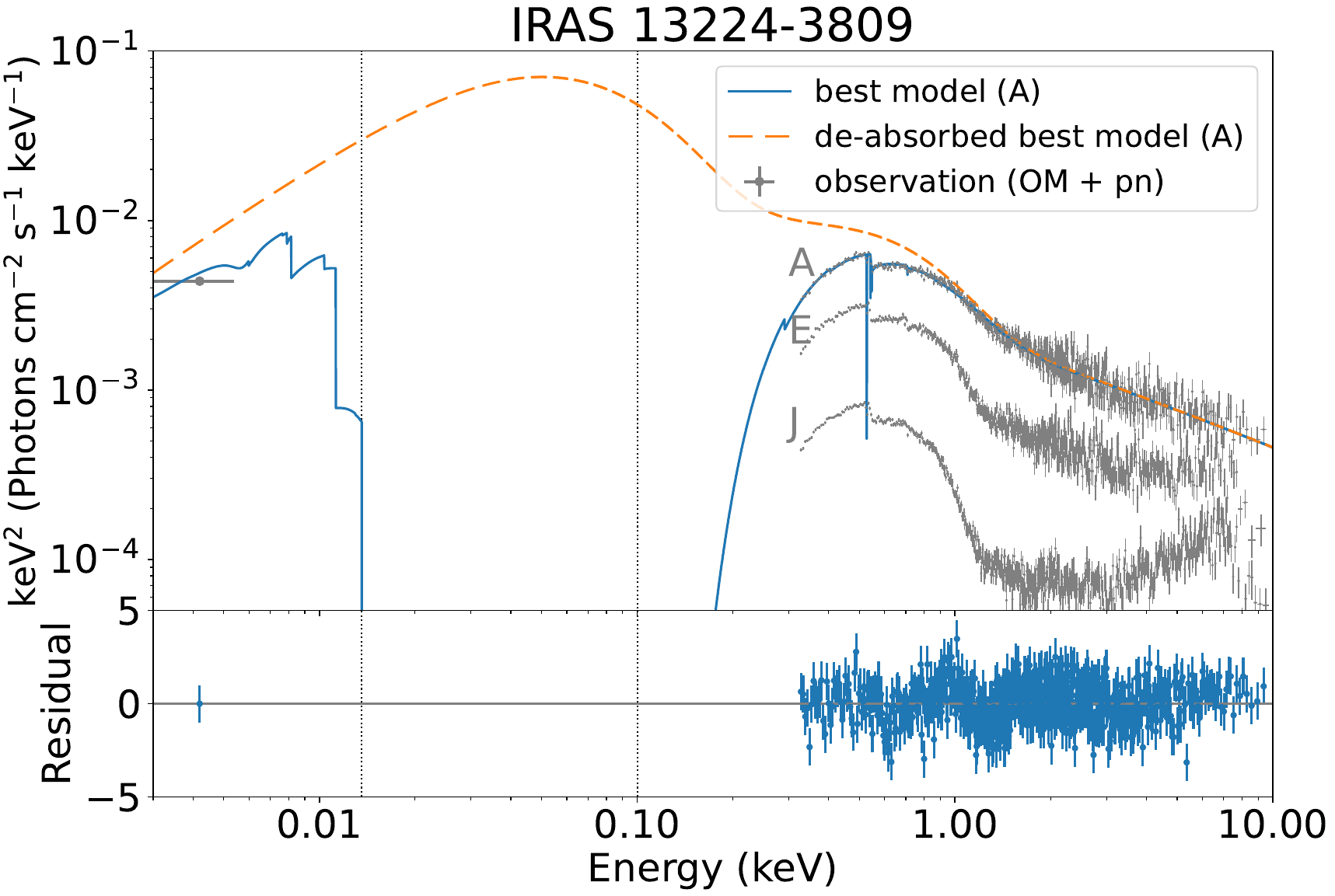}} 
	 \caption{Broadband UV--X-ray SED fitting with pn and OM data in 3~eV--10~keV. Three intensity-sliced spectra A, E, and J are shown in gray. The blue line shows the best-fit model for spectrum A. The orange dashed line shows the deabsorbed best-fit model, from which we can estimate the intrinsic flux in UV (13.6~eV--0.1~keV) and X-rays (0.1--10.0~keV). The vertical dotted lines show the boundary of these energy bands.} 
	    \label{fig:agnslim}
\end{figure}

\begin{table}
    \begin{center}
    \begin{threeparttable}
    \caption{Best-fit parameters determined by the SED fitting obtained by OM and pn.}
    \label{tab:agnslim}
        \begin{tabular}{llc}\hline
        Components & Parameters & Best-fit values \\\hline
        \phabs & \nh (10$^{20}$cm$^{-2}$) & 5 (fixed)  \\
        {\tt redden} & $E(B-V)$ & 0.086 (fixed)  \\
        \agnslim & mass ($M_\odot$)  & 5.4$\times 10^6$ (fixed)  \\
         & $\dot{M}$ ($\dot{M}_{\rm Edd}$) & 6.30$\pm0.01$  \\
         & $a^*$ & 0 (fixed)  \\
         & cos~$i$ &  0.5 (fixed) \\
         & $kT_{\rm hot}$ (keV) & 100 (fixed)  \\
         & $kT_{\rm warm}$ (keV) &  0.166$\pm0.005$ \\
         & $\Gamma_{\rm hot}$ &  2.67$\pm0.04$ \\
         & $\Gamma_{\rm warm}$ &  1.66$^{+0.09}_{-0.10}$ \\
         & $R_{\rm hot}$ ($R_{\rm g}$) &  8.3$\pm0.2$ \\
         & $R_{\rm warm}$ ($R_{\rm g}$) &  10.6$\pm0.1$ \\\hline
	\end{tabular}
    \end{threeparttable}
    \end{center}
\end{table}

\subsection{SED fitting with OM+pn spectra} 
In this section, we model the broadband spectral energy distribution (SED) with the pn and OM data to
estimate the bolometric luminosity,   affecting  the disk winds and ionization degree in the ambient matter.
Since we did not find significant flux variations in the OM data compared to the X-ray variation, we use the time-averaged OM spectrum. 

Since \iras is a super-Eddington object and the AGN component is significant, we ignore the host galaxy contamination.
In this study, we assume that spectrum A is not partially absorbed (CF$=$0), so we model SED of the spectrum A with the \agnslim model \citep{Kubota19}, multiplied by the reddening (\redden; \citealp{Cardelli89}) and the interstellar absorption (\phabs).
Here \agnslim is a broadband spectral model for super-Eddington AGN based on the slim disk emissivity.
The accretion flow is radially stratified; the inner hot Comptonization region ($R_{\rm in}$ to $R_{\rm hot}$) with the electron temperature $kT_{\rm hot}$ and photon index $\Gamma_{\rm hot}$, the intermediate warm Comptonization region to produce the soft X-ray excess ($R_{\rm warm}$ to $R_{\rm hot}$) with the electron temperature $kT_{\rm warm}$ and photon index $\Gamma_{\rm warm}$, and the outer standard disk ($R_{\rm warm}$ to $R_{\rm out}$).

For a highly super-Eddington slim disk, the inner radius of the disk ($R_{\rm in}$) is determined  not only by the black hole spin but also by the gas pressure, so that $R_{\rm in}$ can be smaller than the inner-most stable circular orbit \citep{Watarai00, Kubota19}. We assume the spin parameter $a^*=0$, and set $R_{\rm in}$ to the radius calculated by \cite{Kubota19} (see Equation~1 in their paper).
We also fixed the column density of the interstellar absorption \nh to $5 \times 10^{20}$~cm$^{-2}$ and $E(B-V) = 1.7 \times \nh / 10^{22}$ cm$^{-2}$ \citep{Bohlin78}. 

The averaged OM spectrum and pn spectrum A are successfully fitted with the assumed model.
Figure~\ref{fig:agnslim} shows the $\nu f_{\nu}$ plot of the broadband SED in 3~eV--10~keV. The gray lines indicate the three representative intensity-sliced spectra A, E, and J, and the blue line represents the best-fit SED model for spectrum A.
Best-fit parameters are given in Table~\ref{tab:agnslim}.

\section{Discussion} 
\subsection{Comparison between the UFO and clumpy absorbers} 
\begin{figure} 
\centerline{\includegraphics[width=1.0\columnwidth]{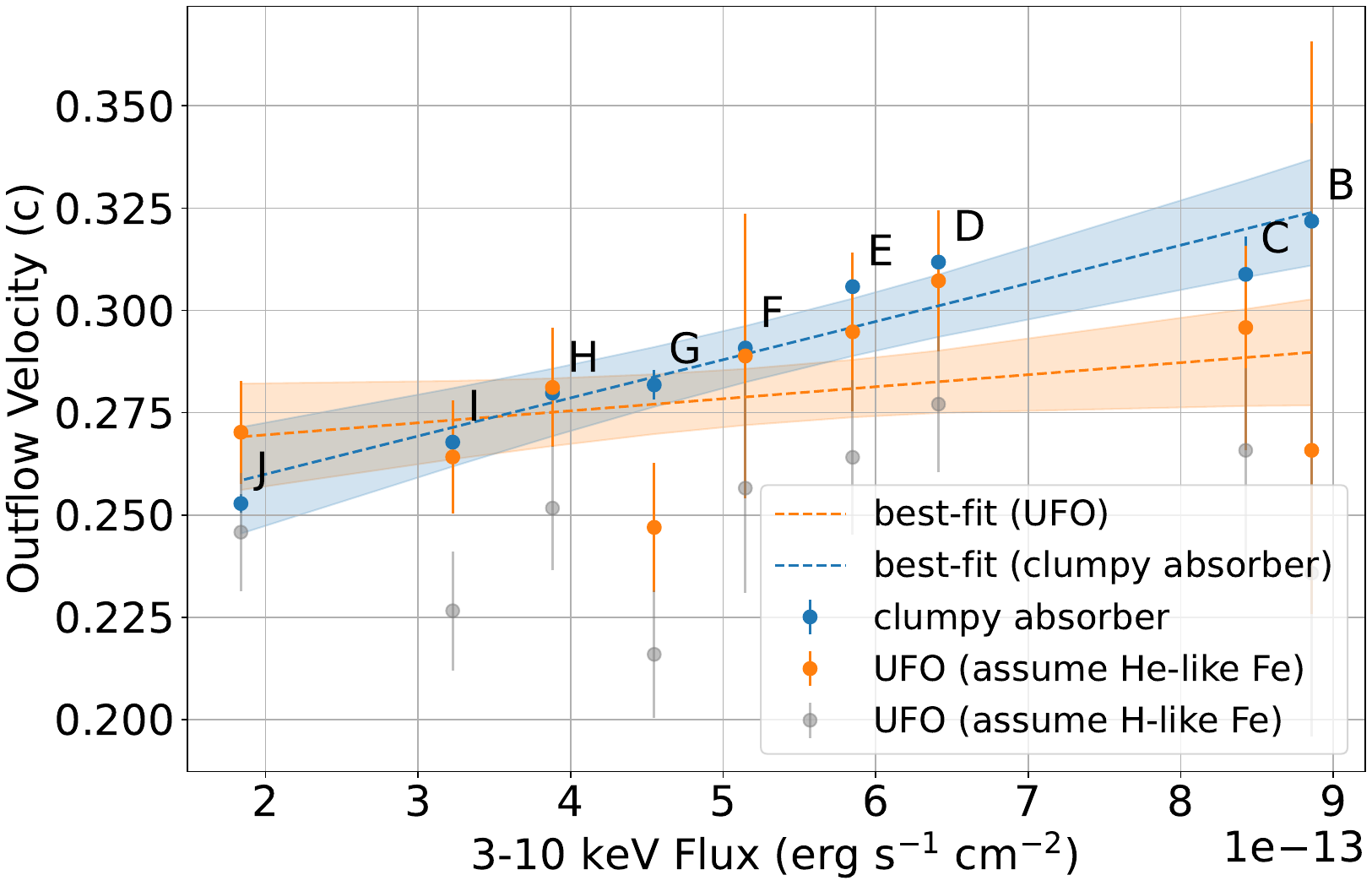}}
	    \caption{Outflow velocities of the clumpy absorbers (in blue) obtained from the spectral ratio fitting, and those of the UFOs estimated from the spectral fitting (in orange and gray assuming He-like and H-like iron, respectively). The outflowing velocities are plotted against the 3-10 keV flux of the intensity-sliced spectra. The blue and orange dotted lines show the best-fit linear functions of the UFOs and clumps with 1$\sigma$ uncertainties.
	    }
	    \label{fig:vel_comp}
\end{figure}

In Section~\ref{sec:3}, we constrained the outflow velocity, \vout, of the partial clumpy absorbers using the unique spectral ratio fitting method. We also performed the conventional spectral fitting to determine the \vout of the UFO.
Comparison of the outflow velocities of the UFOs and partial absorbers are shown in Figure~\ref{fig:vel_comp}, where these velocities are plotted against the 3.0--10.0~keV X-ray flux of the intensity-sliced spectra.
When a UFO is composed of multiple absorption lines with different velocity components, the average velocity weighted by the column density of each absorption line ($N_{\rm atom}$) is plotted. 

We fitted the velocities of the UFOs and clumpy absorbers as a function of the X-ray flux with linear functions.
The orange and blue dotted lines in Figure~\ref{fig:vel_comp} show the best-fit models with 1~$\sigma$ uncertainties.
Although the slopes of the two best-fit functions are slightly different, we see similar increasing trends with the X-ray flux of the clump velocities and the UFO velocities either assuming the He-like or H-like absorption lines.
In particular, the orange dots in Figure~\ref{fig:vel_comp} assuming the He-like Fe absorption show similar velocity values with those of the clumpy absorbers. This finding is consistent with the ``hot inner and clumpy outer wind'' scenario by \cite{Mizumoto19}, where the outer clumpy absorbers are the consequence of the inner hot UFO winds.

\begin{figure} 
\centerline{\includegraphics[width=1.0\columnwidth]{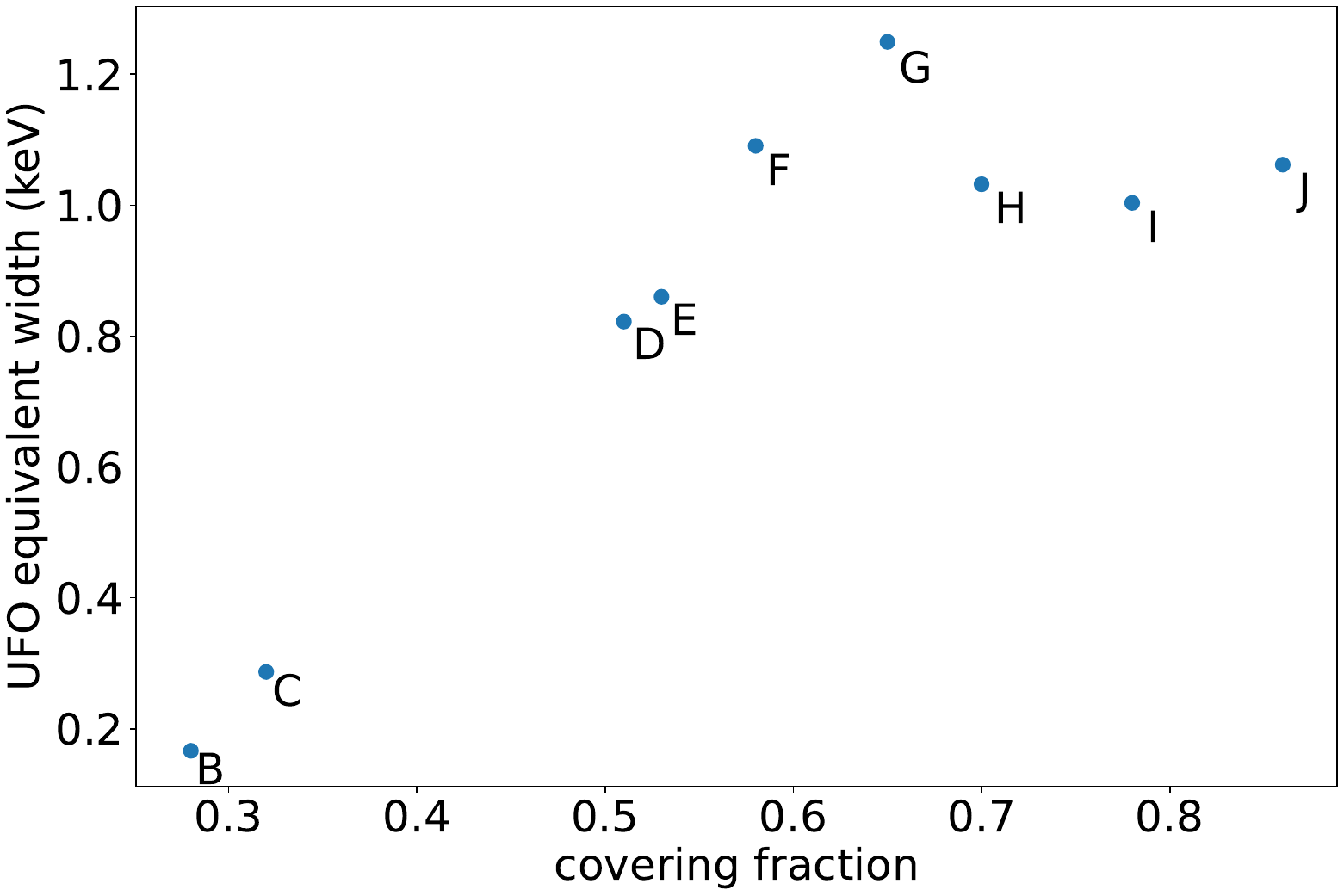}}
	    \caption{The horizontal axis shows the clump covering fraction, and the vertical axis shows the equivalent width of the UFO absorption lines, which is calculated as the sum of individual equivalent widths of multiple absorption lines.}
	    \label{fig:CFvsEW}
\end{figure}

If the proposed scenario is valid, there should be a correlation between the clumps and UFOs along the line of sight. 
To investigate this possibility, we plotted the clump covering fraction (see Table~\ref{tab:ratio_par}) and the equivalent width of the UFO absorption lines (Figure~\ref{fig:CFvsEW}). 
The equivalent width was calculated as the sum of the individual equivalent widths of multiple absorption lines.
Figure~\ref{fig:CFvsEW} indicates a clear correlation between the two variables, in particular a surprising linearity in the B--G range. This finding suggests a simple fact  that more material in the line-of-sight will result in more prominent UFOs and partial covering, providing further evidence of the ``hot inner and clumpy outer wind'' scenario.

These results have also been suggested in PDS~456, which is another super-Eddington source.
\cite{Matzeu16} searched for the outflow velocity of the clumpy absorber to give the best spectral fit, 
and reported the global minimum solution at a velocity of 0.25~$c$.
\cite{Matzeu17a} also found in the same target that the centroid energy of Fe K UFO absorption increases with the ionizing flux. 
\cite{Reeves21} similarly explained the spectrum of PDS~456 in terms of the partial absorption;  
Figure~16 in their paper clearly shows that the UFO absorption lines are more prominent when the soft X-rays  are more significantly absorbed by the partial clumps.

\subsection{Relation of variations between intrinsic UV/X-ray luminosity and UFO velocity} 
As shown in Figure~\ref{fig:sliced}, the observed 0.3 -- 10 keV flux variation ranges over more than one order of magnitude. On the other hand, the ratio fitting results show that the intrinsic X-ray flux is varied by only a factor of $\sim$4, indicating that a considerable amount of the apparent flux variation is caused by changes in the partial covering fraction.

The orange dashed line in Figure~\ref{fig:agnslim} shows the deabsorbed UV-X-ray spectrum for spectrum A in which the covering fraction is assumed to be null. 
From this deabsorbed model, the intrinsic UV flux in 13.6~eV -- 0.1~keV and the intrinsic X-ray flux in 0.1--10.0~keV are estimated respectively to be $1.8 \times 10^{-10}$ erg~cm$^{-2}$~s$^{-1}$ and $6.1\times 10^{-11}$ erg~cm$^{-2}$~s$^{-1}$, using {\tt flux} command in XSPEC.
The proportion of the intrinsic UV flux to X-ray flux is approximately $3:1$.
Since the intrinsic X-ray flux varies by a factor of $\sim$4 while the UV flux is invariable, the total flux variation of UV and X-rays can be estimated to be about a factor of $\sim$16/13, assuming that the obtained ratio (UV flux: X-ray flux = 3:1) is valid throughout the observation period.

In case the disk wind is driven by continuum radiation, the total flux variation is expected to contribute directly to the increase in the kinetic energy of the disk wind. Since the outflow velocity is proportional to the square root of the wind kinetic energy, the velocity change is estimated to be about a factor of $\sqrt{16/13} \sim$1.1. 

By extrapolating the best-fit model in Figure~\ref{fig:vel_comp}, the velocities of the UFO and clumpy absorbers in spectrum A ($1.4 \times 10^{-12}$ erg~cm$^{-2}$~s$^{-1}$ in 3--10~keV) are estimated to be 0.31~$c$ and 0.37~$c$.
As a result, the variations of the UFO and clump velocities from the dimmest spectrum (J) to the brightest one (A) are about a factor of 1.1 and 1.4, respectively.
These values are consistent with the simple estimation above, thus both the clumpy absorbers and the UFOs are considered to be driven by the same UV-dominant continuum radiation.

\subsection{Constant ionization parameter of the clumps} 
Ionization parameter ($\xi = L/n r^2$) values of the clumpy absorbers determined from the model fitting are hardly variable regardless of the X-ray flux (Table \ref{tab:ratio_par} and \ref{tab:best_specfit}). 
This fact suggests that the gas density of the clump, $n$, increases or that the location where the clump is formed, $r$, gets more distant as the X-ray luminosity increases. 

\cite{Takeuchi14} suggests that the clump is formed by a sort of radiation hydrodynamic instability in addition to the Rayleigh-Taylor instability. 
The radiation instability should work only at a certain opacity, i.e., a certain $\xi$. If the instability is activated and the clumps are produced at this specific distance which increases with the luminosity, the result that $\xi$ being constant is reasonable.
This may be a common feature of the super-Eddington objects with continuum-driven outflows.
It will be intriguing to investigate the flux dependence of $\xi$ in other super-Eddington sources in future.

\subsection{Explanation of other observed properties}
\iras has been observed many times and extensively investigated to date.
This section will discuss how our model can explain these observed properties of \iras.

In our model, spectral variability is explained  by the variation of the PL continuum component and the partial absorption component. 
On the other hand, \cite{Parker17} used the principle component analysis (PCA) to decompose the spectral variability, and argued that the PL emission dominates the spectral variability, with a minor contribution from the  reflection and   UFO component.
The PCA is a powerful data science tool to reduce the dimensionality of the input data, but one must be careful with its applicability in X-ray spectral analysis. 
\cite{Pike17} performed a spectral component decomposition of the simulated X-ray spectral variations of NLS1 galaxies using the PCA and the non-negative matrix factorization (NMF).
Despite the PCA's good performance in determining the number of the independent spectral components, the decomposed components derived by the PCA include the  {\em negative}\/ values as residuals from the average spectrum, being  very different from the original components.
The other method, NMF, has the advantage of being composed of non-negative values, making it easier to interpret the spectral components, in particular when the variable components are {\em additive}.  
Still, NMF is not very useful when variable spectral components are {\em multiplicative}, just like  our model.
We expect that forthcoming analyses using new techniques will reveal the main cause of the spectral variation without prior assumptions.

X-ray observations provide not only spectroscopic information but also timing information.
Characteristic power spectral density (PSD) and soft/hard lags have been reported in \iras.
The PSD peaks at timescales of about 30~ks (e.g., \citealp{Alston19}).
The radiation-magnetohydrodynamic simulations suggest the clumpy absorbers to exist at a few hundred R$_{\rm s}$ from the center \citep{Takeuchi13}.
Assuming the clump exists at 500~R$_{\rm s}$, the Keplerian rotation period is calculated to be about 200~ks.
Since the typical clump size is around 10~R$_{\rm s}$  \citep{Takeuchi13}, a  variation of 30~ks is reasonable with our model in terms of changes in  the partial covering fraction.

The ``lag-frequency plot'' can be derived from the light curves of the primary X-ray emission (typically 1.0--4.0~keV) and the soft excess (below $\sim$1.0~keV). 
In this plot, a hard lag is observed in the low-frequency band below 10$^{-4}$~Hz, while a soft lag is observed in the high-frequency band above 10$^{-3}$~Hz (e.g., \citealp{DeMarco13} and \citealp{Alston20}).
The low-frequency hard lag has a PL-like shape and decreases with frequency.
This is considered to be due to inward propagation of the fluctuations through the accretion flow and corona (e.g., \citealp{Arevalo06}).

Since the mechanism of the soft X-ray excess has not yet been clearly understood, two major hypotheses exist as to the cause of the high-frequency soft lag.
One is reverberation due to reflection by the inner region of the accretion disk around $\sim$10~R$_{\rm s}$ (e.g., \citealp{Fabian09} and \citealp{Zoghbi12}), where the reprocessed emission is delayed by the light travel time from the source.
The other hypothesis says that the lag is a solely mathematical artifact due to phase wrapping \citep{Miller10a, Legg12, Mizumoto19}, which is also favored by our model.

When examining the energy dependence of the lag in the high-frequency range where the soft lag is observed, a significant peak structure is often  found in the Fe K band. 
This is thought to be caused by the primary X-rays scattering off the surrounding material, generating the Fe emission lines with a reverberation lag (e.g., \citealp{Zoghbi12} and \citealp{Kara13}).
If this lag is literally taken as the light propagation time due to the reflection in the inner accretion disk, the size is about 10~R$_{\rm s}$.
On the other hand, if we take into account the dilution effect due to the weaker iron emission line compared to the continuum, the lag can be attributed to the reflections from the materials located as far away as $\sim$100~R$_{\rm s}$ (e.g., \citealp{Miller10}).
We speculate that the scattering by the disk wind in our model at $\sim$100~R$_{\rm s}$ can explain the observed reverberation lag \citep{Mizumoto18a, Mizumoto19}.

X-ray light curves of almost any kinds of accreting objects, including AGNs, exhibit a linear relationship between the root mean square (RMS) amplitude and the flux variations (e.g., \citealp{Uttley01} and \citealp{Alston19}).
The fact that this relationship has been observed for any objects with accretion disks suggests that it is due to intrinsic X-ray variability, not due to apparent absorption effects.
We suppose that the intrinsic PL variability in our model is responsible for this relation, while quantitative estimation is a future subject.

\section{Conclusion}\label{sec:5} 
We have applied the unique ``spectral-ratio model fitting'' technique for the first time to disentangle 
spectral model/parameter degeneracy of the NLS1 \iras. Taking the spectral ratios of the intensity-sliced spectra allows us to focus on variable absorption features by canceling the time-invariable continuum and other absorption components.
Consequently, we found that the soft X-ray spectral variation is mostly explained by the change in the covering fraction of the mildly-ionized clumpy absorbers, and that the clumps are outflowing at noticeably high velocities comparable to those of UFOs ($\sim$0.2--0.3~$c$).

We found that both velocities of the partially absorbing clumps and the UFOs increase with the intrinsic UV-X-ray luminosity, and the luminosity dependence suggests that the outflowing velocity being consistent with the kinetic velocity of the radiatively driven winds.
This is the first observational evidence that the clumps responsible for the partial absorption have the same origin as the radiation-driven UFOs.
Present results support the ``hot inner and clumpy outer wind'' scenario \citep{Mizumoto19}, where the clumpy absorbers are originated from the inner hot UFO winds driven by the UV-dominant continuum radiation.

\begin{acknowledgments}
This research has used data and software provided by the High Energy Astrophysics Science Archive Research Center (HEASARC), a service of the Astrophysics Science Division at NASA/GSFC and the High Energy Astrophysics Division of the Smithsonian Astrophysical Observatory. This study was based on observations obtained with \xmm, ESA science missions with instruments and contributions directly funded by ESA Member States and NASA. This research was supported by JSPS Grant-in-Aid for JSPS Research Fellow Grant Number JP20J20809 (T.M.), JSPS KAKENHI Grant Number JP21K13958 (M.M). M.M. acknowledges support from the Hakubi project at Kyoto University.
\end{acknowledgments}

%

\vspace{5mm}
\facilities{XMM-Newton(pn, OM)}

\software{XSTAR \citep{Kallman01, Kallman04}, XSPEC \citep{Arnaud96}, SAS \citep{Gabriel04}} 

\bibliography{main}

\begin{thebibliography}{}
\expandafter\ifx\csname natexlab\endcsname\relax\def\natexlab#1{#1}\fi
\providecommand{\url}[1]{\href{#1}{#1}}
\providecommand{\dodoi}[1]{doi:~\href{http://doi.org/#1}{\nolinkurl{#1}}}
\providecommand{\doeprint}[1]{\href{http://ascl.net/#1}{\nolinkurl{http://ascl.net/#1}}}
\providecommand{\doarXiv}[1]{\href{https://arxiv.org/abs/#1}{\nolinkurl{https://arxiv.org/abs/#1}}}

\bibitem[{Alston {et~al.}(2019)Alston, Fabian, Buisson, Kara, Parker, Lohfink,
  Uttley, Wilkins, Pinto, De~Marco, Cackett, Middleton, Walton, Reynolds,
  Jiang, Gallo, Zogbhi, Miniutti, Dovciak, \& Young}]{Alston19}
Alston, W.~N., Fabian, A.~C., Buisson, D. J.~K., {et~al.} 2019, Monthly Notices
  of the Royal Astronomical Society, 482, 2088, \dodoi{10.1093/mnras/sty2527}

\bibitem[{Alston {et~al.}(2020)Alston, Fabian, Kara, Parker, Dovciak, Pinto,
  Jiang, Middleton, Miniutti, Walton, Wilkins, Buisson, {Caballero-Garcia},
  Cackett, De~Marco, Gallo, Lohfink, Reynolds, Uttley, Young, \&
  Zogbhi}]{Alston20}
Alston, W.~N., Fabian, A.~C., Kara, E., {et~al.} 2020, Nature Astronomy, 4,
  597, \dodoi{10.1038/s41550-019-1002-x}

\bibitem[{Ar{\'e}valo \& Uttley(2006)}]{Arevalo06}
Ar{\'e}valo, P., \& Uttley, P. 2006, Monthly Notices of the Royal Astronomical
  Society, 367, 801, \dodoi{10.1111/j.1365-2966.2006.09989.x}

\bibitem[{Arnaud(1996)}]{Arnaud96}
Arnaud, K.~A. 1996, Astronomical Data Analysis Software and Systems, 101, 17

\bibitem[{Bekhti {et~al.}(2016)Bekhti, Fl{\"o}er, Keller, Kerp, Lenz, Winkel,
  Bailin, Calabretta, Dedes, Ford, Gibson, Haud, Janowiecki, Kalberla, Lockman,
  {McClure-Griffiths}, Murphy, Nakanishi, Pisano, \&
  {Staveley-Smith}}]{Bekhti16}
Bekhti, N.~B., Fl{\"o}er, L., Keller, R., {et~al.} 2016, Astronomy \&
  Astrophysics, 594, A116, \dodoi{10.1051/0004-6361/201629178}

\bibitem[{Bohlin {et~al.}(1978)Bohlin, Savage, \& Drake}]{Bohlin78}
Bohlin, R.~C., Savage, B.~D., \& Drake, J.~F. 1978, The Astrophysical Journal,
  224, 132, \dodoi{10.1086/156357}

\bibitem[{Cardelli {et~al.}(1989)Cardelli, Clayton, \& Mathis}]{Cardelli89}
Cardelli, J.~A., Clayton, G.~C., \& Mathis, J.~S. 1989, The Astrophysical
  Journal, 345, 245, \dodoi{10.1086/167900}

\bibitem[{Chartas \& Canas(2018)}]{Chartas18}
Chartas, G., \& Canas, M.~H. 2018, The Astrophysical Journal, 867, 103,
  \dodoi{10.3847/1538-4357/aae438}

\bibitem[{Chiang {et~al.}(2015)Chiang, Walton, Fabian, Wilkins, \&
  Gallo}]{Chiang15}
Chiang, C.-Y., Walton, D.~J., Fabian, A.~C., Wilkins, D.~R., \& Gallo, L.~C.
  2015, Monthly Notices of the Royal Astronomical Society, 446, 759,
  \dodoi{10.1093/mnras/stu2087}

\bibitem[{Dannen {et~al.}(2020)Dannen, Proga, Waters, \& Dyda}]{Dannen20}
Dannen, R.~C., Proga, D., Waters, T., \& Dyda, S. 2020, The Astrophysical
  Journal, 893, L34, \dodoi{10.3847/2041-8213/ab87a5}

\bibitem[{De~Marco {et~al.}(2013)De~Marco, Ponti, Cappi, Dadina, Uttley,
  Cackett, Fabian, \& Miniutti}]{DeMarco13}
De~Marco, B., Ponti, G., Cappi, M., {et~al.} 2013, Monthly Notices of the Royal
  Astronomical Society, 431, 2441, \dodoi{10.1093/mnras/stt339}

\bibitem[{Di~Gesu {et~al.}(2015)Di~Gesu, Costantini, Ebrero, Mehdipour,
  Kaastra, Ursini, Petrucci, Cappi, Kriss, Bianchi, {Branduardi-Raymont},
  De~Marco, De~Rosa, Kaspi, Paltani, Pinto, Ponti, Steenbrugge, \&
  Whewell}]{DiGesu15}
Di~Gesu, L., Costantini, E., Ebrero, J., {et~al.} 2015, Astronomy and
  Astrophysics, 579, A42, \dodoi{10.1051/0004-6361/201525934}

\bibitem[{Emmanoulopoulos {et~al.}(2014)Emmanoulopoulos, Papadakis, Dov{\v
  c}iak, \& McHardy}]{Emmanoulopoulos14}
Emmanoulopoulos, D., Papadakis, I.~E., Dov{\v c}iak, M., \& McHardy, I.~M.
  2014, Monthly Notices of the Royal Astronomical Society, 439, 3931,
  \dodoi{10.1093/mnras/stu249}

\bibitem[{Fabian {et~al.}(1989)Fabian, Rees, Stella, \& White}]{Fabian89}
Fabian, A.~C., Rees, M.~J., Stella, L., \& White, N.~E. 1989, Monthly Notices
  of the Royal Astronomical Society, 238, 729, \dodoi{10.1093/mnras/238.3.729}

\bibitem[{Fabian {et~al.}(2009)Fabian, Zoghbi, Ross, Uttley, Gallo, Brandt,
  Blustin, Boller, {Caballero-Garcia}, Larsson, Miller, Miniutti, Ponti, Reis,
  Reynolds, Tanaka, \& Young}]{Fabian09}
Fabian, A.~C., Zoghbi, A., Ross, R.~R., {et~al.} 2009, Nature, 459, 540,
  \dodoi{10.1038/nature08007}

\bibitem[{Fabian {et~al.}(2013)Fabian, Kara, Walton, Wilkins, Ross, Lozanov,
  Uttley, Gallo, Zoghbi, Miniutti, Boller, Brandt, Cackett, Chiang, Dwelly,
  Malzac, Miller, Nardini, Ponti, Reis, Reynolds, Steiner, Tanaka, \&
  Young}]{Fabian13}
Fabian, A.~C., Kara, E., Walton, D.~J., {et~al.} 2013, Monthly Notices of the
  Royal Astronomical Society, 429, 2917, \dodoi{10.1093/mnras/sts504}

\bibitem[{{Foreman-Mackey} {et~al.}(2013){Foreman-Mackey}, Hogg, Lang, \&
  Goodman}]{Foreman-Mackey13}
{Foreman-Mackey}, D., Hogg, D.~W., Lang, D., \& Goodman, J. 2013, Publications
  of the Astronomical Society of the Pacific, 125, 306, \dodoi{10.1086/670067}

\bibitem[{Gabriel {et~al.}(2004)Gabriel, Denby, Fyfe, Hoar, Ibarra, Ojero,
  Osborne, Saxton, Lammers, \& Vacanti}]{Gabriel04}
Gabriel, C., Denby, M., Fyfe, D.~J., {et~al.} 2004, 314, 759

\bibitem[{Gofford {et~al.}(2013)Gofford, Reeves, Tombesi, Braito, Turner,
  Miller, \& Cappi}]{Gofford13}
Gofford, J., Reeves, J.~N., Tombesi, F., {et~al.} 2013, Monthly Notices of the
  Royal Astronomical Society, 430, 60, \dodoi{10.1093/mnras/sts481}

\bibitem[{Hagino {et~al.}(2015)Hagino, Odaka, Done, Gandhi, Watanabe, Sako, \&
  Takahashi}]{Hagino15}
Hagino, K., Odaka, H., Done, C., {et~al.} 2015, Monthly Notices of the Royal
  Astronomical Society, 446, 663, \dodoi{10.1093/mnras/stu2095}

\bibitem[{Igo {et~al.}(2020)Igo, Parker, Matzeu, Alston, Alvarez~Crespo,
  F{\"u}rst, Buisson, Lobban, Joyce, Mallick, Schartel, \&
  {Santos-Lle{\'o}}}]{Igo20}
Igo, Z., Parker, M.~L., Matzeu, G.~A., {et~al.} 2020, Monthly Notices of the
  Royal Astronomical Society, 493, 1088, \dodoi{10.1093/mnras/staa265}

\bibitem[{Jansen {et~al.}(2001)Jansen, Lumb, Altieri, Clavel, Ehle, Erd,
  Gabriel, Guainazzi, Gondoin, Much, Munoz, Santos, Schartel, Texier, \&
  Vacanti}]{Jansen01}
Jansen, F., Lumb, D., Altieri, B., {et~al.} 2001, Astronomy \& Astrophysics,
  365, L1, \dodoi{10.1051/0004-6361:20000036}

\bibitem[{Jiang {et~al.}(2022)Jiang, Dauser, Fabian, Alston, Gallo, Parker, \&
  Reynolds}]{Jiang22a}
Jiang, J., Dauser, T., Fabian, A.~C., {et~al.} 2022, arXiv:2204.09908
  [astro-ph].
\newblock \doeprint{2204.09908}

\bibitem[{Kaastra {et~al.}(2014)Kaastra, Kriss, Cappi, Mehdipour, Petrucci,
  Steenbrugge, Arav, Behar, Bianchi, Boissay, {Branduardi-Raymont},
  Chamberlain, Costantini, Ely, Ebrero, Di~Gesu, Harrison, Kaspi, Malzac,
  De~Marco, Matt, Nandra, Paltani, Person, Peterson, Pinto, Ponti, Nu{\~n}ez,
  De~Rosa, Seta, Ursini, {de Vries}, Walton, \& Whewell}]{Kaastra14}
Kaastra, J.~S., Kriss, G.~A., Cappi, M., {et~al.} 2014, Science, 345, 64,
  \dodoi{10.1126/science.1253787}

\bibitem[{Kallman \& Bautista(2001)}]{Kallman01}
Kallman, T., \& Bautista, M. 2001, The Astrophysical Journal Supplement Series,
  133, 221, \dodoi{10.1086/319184}

\bibitem[{Kallman {et~al.}(2004)Kallman, Palmeri, Bautista, Mendoza, \&
  Krolik}]{Kallman04}
Kallman, T.~R., Palmeri, P., Bautista, M.~A., Mendoza, C., \& Krolik, J.~H.
  2004, The Astrophysical Journal Supplement Series, 155, 675,
  \dodoi{10.1086/424039}

\bibitem[{Kara {et~al.}(2013)Kara, Fabian, Cackett, Steiner, Uttley, Wilkins,
  \& Zoghbi}]{Kara13}
Kara, E., Fabian, A.~C., Cackett, E.~M., {et~al.} 2013, Monthly Notices of the
  Royal Astronomical Society, 428, 2795, \dodoi{10.1093/mnras/sts155}

\bibitem[{King(2010)}]{King10}
King, A. 2010, Monthly Notices of the Royal Astronomical Society, 402, 1516,
  \dodoi{10.1111/j.1365-2966.2009.16013.x}

\bibitem[{King \& Pounds(2015)}]{King15}
King, A., \& Pounds, K. 2015, Annual Review of Astronomy and Astrophysics, 53,
  115, \dodoi{10.1146/annurev-astro-082214-122316}

\bibitem[{Kobayashi {et~al.}(2018)Kobayashi, Ohsuga, Takahashi, Kawashima,
  Asahina, Takeuchi, \& Mineshige}]{Kobayashi18}
Kobayashi, H., Ohsuga, K., Takahashi, H.~R., {et~al.} 2018, Publications of the
  Astronomical Society of Japan, 70, 22, \dodoi{10.1093/pasj/psx157}

\bibitem[{Kubota \& Done(2019)}]{Kubota19}
Kubota, A., \& Done, C. 2019, Monthly Notices of the Royal Astronomical
  Society, 489, 524, \dodoi{10.1093/mnras/stz2140}

\bibitem[{Laha {et~al.}(2014)Laha, Guainazzi, Dewangan, Chakravorty, \&
  Kembhavi}]{Laha14}
Laha, S., Guainazzi, M., Dewangan, G.~C., Chakravorty, S., \& Kembhavi, A.~K.
  2014, Monthly Notices of the Royal Astronomical Society, 441, 2613,
  \dodoi{10.1093/mnras/stu669}

\bibitem[{Laha {et~al.}(2021)Laha, Reynolds, Reeves, Kriss, Guainazzi, Smith,
  Veilleux, \& Proga}]{Laha21}
Laha, S., Reynolds, C.~S., Reeves, J., {et~al.} 2021, Nature Astronomy, 5, 13,
  \dodoi{10.1038/s41550-020-01255-2}

\bibitem[{Legg {et~al.}(2012)Legg, Miller, Turner, Giustini, Reeves, \&
  Kraemer}]{Legg12}
Legg, E., Miller, L., Turner, T.~J., {et~al.} 2012, The Astrophysical Journal,
  760, 73, \dodoi{10.1088/0004-637X/760/1/73}

\bibitem[{Mason {et~al.}(2001)Mason, Breeveld, Much, Carter, Cordova, Cropper,
  Fordham, Huckle, Ho, Kawakami, Kennea, Kennedy, Mittaz, Pandel, Priedhorsky,
  Sasseen, Shirey, Smith, \& Vreux}]{Mason01}
Mason, K.~O., Breeveld, A., Much, R., {et~al.} 2001, Astronomy \& Astrophysics,
  365, L36, \dodoi{10.1051/0004-6361:20000044}

\bibitem[{Matzeu {et~al.}(2017)Matzeu, Reeves, Braito, Nardini, McLaughlin,
  Lobban, Tombesi, \& Costa}]{Matzeu17a}
Matzeu, G.~A., Reeves, J.~N., Braito, V., {et~al.} 2017, Monthly Notices of the
  Royal Astronomical Society: Letters, 472, L15, \dodoi{10.1093/mnrasl/slx129}

\bibitem[{Matzeu {et~al.}(2016)Matzeu, Reeves, Nardini, Braito, Costa, Tombesi,
  \& Gofford}]{Matzeu16}
Matzeu, G.~A., Reeves, J.~N., Nardini, E., {et~al.} 2016, Monthly Notices of
  the Royal Astronomical Society, 458, 1311, \dodoi{10.1093/mnras/stw354}

\bibitem[{Matzeu {et~al.}(2023)Matzeu, Brusa, Lanzuisi, Dadina, Bianchi, Kriss,
  Mehdipour, Nardini, Chartas, Middei, Piconcelli, Gianolli, Comastri,
  Longinotti, Krongold, Ricci, Petrucci, Tombesi, Luminari, Zappacosta,
  Miniutti, Gaspari, Behar, Bischetti, Mathur, Perna, Giustini, Grandi,
  Torresi, Vignali, Bruni, Cappi, Costantini, Cresci, Marco, Rosa, Gilli,
  Guainazzi, Kaastra, Kraemer, Franca, Marconi, Panessa, Ponti, Proga, Ursini,
  Baldini, Fiore, King, Maiolino, Matt, \& Merloni}]{Matzeu23}
Matzeu, G.~A., Brusa, M., Lanzuisi, G., {et~al.} 2023, Astronomy \&
  Astrophysics, 670, A182, \dodoi{10.1051/0004-6361/202245036}

\bibitem[{McKernan {et~al.}(2007)McKernan, Yaqoob, \& Reynolds}]{McKernan07}
McKernan, B., Yaqoob, T., \& Reynolds, C.~S. 2007, Monthly Notices of the Royal
  Astronomical Society, 379, 1359, \dodoi{10.1111/j.1365-2966.2007.11993.x}

\bibitem[{Midooka {et~al.}(2022{\natexlab{a}})Midooka, Ebisawa, Mizumoto, \&
  Sugawara}]{Midooka22}
Midooka, T., Ebisawa, K., Mizumoto, M., \& Sugawara, Y. 2022{\natexlab{a}},
  Monthly Notices of the Royal Astronomical Society, stac1206,
  \dodoi{10.1093/mnras/stac1206}

\bibitem[{Midooka {et~al.}(2022{\natexlab{b}})Midooka, Mizumoto, \&
  Ebisawa}]{Midooka22a}
Midooka, T., Mizumoto, M., \& Ebisawa, K. 2022{\natexlab{b}}

\bibitem[{Miller {et~al.}(2010{\natexlab{a}})Miller, Turner, Reeves, \&
  Braito}]{Miller10a}
Miller, L., Turner, T.~J., Reeves, J.~N., \& Braito, V. 2010{\natexlab{a}},
  Monthly Notices of the Royal Astronomical Society, 408, 1928,
  \dodoi{10.1111/j.1365-2966.2010.17261.x}

\bibitem[{Miller {et~al.}(2010{\natexlab{b}})Miller, Turner, Reeves, Lobban,
  Kraemer, \& Crenshaw}]{Miller10}
Miller, L., Turner, T.~J., Reeves, J.~N., {et~al.} 2010{\natexlab{b}}, Monthly
  Notices of the Royal Astronomical Society, 403, 196,
  \dodoi{10.1111/j.1365-2966.2009.16149.x}

\bibitem[{Mizumoto {et~al.}(2018)Mizumoto, Done, Hagino, Ebisawa, Tsujimoto, \&
  Odaka}]{Mizumoto18a}
Mizumoto, M., Done, C., Hagino, K., {et~al.} 2018, Monthly Notices of the Royal
  Astronomical Society, 478, 971, \dodoi{10.1093/mnras/sty1114}

\bibitem[{Mizumoto {et~al.}(2019)Mizumoto, Ebisawa, Tsujimoto, Done, Hagino, \&
  Odaka}]{Mizumoto19}
Mizumoto, M., Ebisawa, K., Tsujimoto, M., {et~al.} 2019, Monthly Notices of the
  Royal Astronomical Society, 482, 5316, \dodoi{10.1093/mnras/sty3056}

\bibitem[{Nardini {et~al.}(2015)Nardini, Reeves, Gofford, Harrison, Risaliti,
  Braito, Costa, Matzeu, Walton, Behar, Boggs, Christensen, Craig, Hailey,
  Matt, Miller, O'Brien, Stern, Turner, \& Ward}]{Nardini15}
Nardini, E., Reeves, J.~N., Gofford, J., {et~al.} 2015, Science, 347, 860,
  \dodoi{10.1126/science.1259202}

\bibitem[{Parker {et~al.}(2022)Parker, Matzeu, Matthews, Middleton, Dauser,
  Jiang, \& Joyce}]{Parker22}
Parker, M.~L., Matzeu, G.~A., Matthews, J.~H., {et~al.} 2022, Monthly Notices
  of the Royal Astronomical Society, 513, 551, \dodoi{10.1093/mnras/stac877}

\bibitem[{Parker {et~al.}(2017)Parker, Pinto, Fabian, Lohfink, Buisson, Alston,
  Kara, Cackett, Chiang, Dauser, De~Marco, Gallo, Garcia, Harrison, King,
  Middleton, Miller, Miniutti, Reynolds, Uttley, Vasudevan, Walton, Wilkins, \&
  Zoghbi}]{Parker17}
Parker, M.~L., Pinto, C., Fabian, A.~C., {et~al.} 2017, Nature, 543, 83,
  \dodoi{10.1038/nature21385}

\bibitem[{Pike {et~al.}(2017)Pike, Ebisawa, Ikeda, Morii, Mizumoto, \&
  Kusunoki}]{Pike17}
Pike, S., Ebisawa, K., Ikeda, S., {et~al.} 2017, Journal of Space Science
  Informatics Japan, 6, 73,
  \dodoi{https://doi.org/10.20637/JAXA-RR-16-007/0007}

\bibitem[{Pinto {et~al.}(2018)Pinto, Alston, Parker, Fabian, Gallo, Buisson,
  Walton, Kara, Jiang, Lohfink, \& Reynolds}]{Pinto18}
Pinto, C., Alston, W., Parker, M.~L., {et~al.} 2018, Monthly Notices of the
  Royal Astronomical Society, 476, 1021, \dodoi{10.1093/mnras/sty231}

\bibitem[{Reeves {et~al.}(2008)Reeves, Done, Pounds, Terashima, Hayashida,
  Anabuki, Uchino, \& Turner}]{Reeves08}
Reeves, J., Done, C., Pounds, K., {et~al.} 2008, Monthly Notices of the Royal
  Astronomical Society, 385, L108, \dodoi{10.1111/j.1745-3933.2008.00443.x}

\bibitem[{Reeves {et~al.}(2021)Reeves, Braito, Porquet, Lobban, Matzeu, \&
  Nardini}]{Reeves21}
Reeves, J.~N., Braito, V., Porquet, D., {et~al.} 2021, Monthly Notices of the
  Royal Astronomical Society, 500, 1974, \dodoi{10.1093/mnras/staa3377}

\bibitem[{Str{\"u}der {et~al.}(2001)Str{\"u}der, Briel, Dennerl, Hartmann,
  Kendziorra, Meidinger, Pfeffermann, Reppin, Aschenbach, Bornemann,
  Br{\"a}uninger, Burkert, Elender, Freyberg, Haberl, Hartner, Heuschmann,
  Hippmann, Kastelic, Kemmer, Kettenring, Kink, Krause, M{\"u}ller, Oppitz,
  Pietsch, Popp, Predehl, Read, Stephan, St{\"o}tter, Tr{\"u}mper, Holl,
  Kemmer, Soltau, St{\"o}tter, Weber, Weichert, {von Zanthier}, Carathanassis,
  Lutz, Richter, Solc, B{\"o}ttcher, Kuster, Staubert, Abbey, Holland, Turner,
  Balasini, Bignami, La~Palombara, Villa, Buttler, Gianini, Lain{\'e}, Lumb, \&
  Dhez}]{Struder01}
Str{\"u}der, L., Briel, U., Dennerl, K., {et~al.} 2001, Astronomy \&
  Astrophysics, 365, L18, \dodoi{10.1051/0004-6361:20000066}

\bibitem[{Takeuchi {et~al.}(2013)Takeuchi, Ohsuga, \& Mineshige}]{Takeuchi13}
Takeuchi, S., Ohsuga, K., \& Mineshige, S. 2013, Publications of the
  Astronomical Society of Japan, 65, 88, \dodoi{10.1093/pasj/65.4.88}

\bibitem[{Takeuchi {et~al.}(2014)Takeuchi, Ohsuga, \& Mineshige}]{Takeuchi14}
---. 2014, Publications of the Astronomical Society of Japan, 66, 48,
  \dodoi{10.1093/pasj/psu011}

\bibitem[{Tomaru {et~al.}(2020)Tomaru, Done, Ohsuga, Odaka, \&
  Takahashi}]{Tomaru20}
Tomaru, R., Done, C., Ohsuga, K., Odaka, H., \& Takahashi, T. 2020, Monthly
  Notices of the Royal Astronomical Society, 494, 3413,
  \dodoi{10.1093/mnras/staa961}

\bibitem[{Tombesi {et~al.}(2010)Tombesi, Cappi, Reeves, Palumbo, Yaqoob,
  Braito, \& Dadina}]{Tombesi10a}
Tombesi, F., Cappi, M., Reeves, J.~N., {et~al.} 2010, Astronomy and
  Astrophysics, 521, A57, \dodoi{10.1051/0004-6361/200913440}

\bibitem[{Ueda {et~al.}(2004)Ueda, Murakami, Yamaoka, Dotani, \&
  Ebisawa}]{Ueda04}
Ueda, Y., Murakami, H., Yamaoka, K., Dotani, T., \& Ebisawa, K. 2004, The
  Astrophysical Journal, 609, 325, \dodoi{10.1086/420973}

\bibitem[{Uttley \& McHardy(2001)}]{Uttley01}
Uttley, P., \& McHardy, I.~M. 2001, Monthly Notices of the Royal Astronomical
  Society, 323, L26, \dodoi{10.1046/j.1365-8711.2001.04496.x}

\bibitem[{Watarai {et~al.}(2000)Watarai, Fukue, Takeuchi, \&
  Mineshige}]{Watarai00}
Watarai, K.-y., Fukue, J., Takeuchi, M., \& Mineshige, S. 2000, Publications of
  the Astronomical Society of Japan, 52, 133, \dodoi{10.1093/pasj/52.1.133}

\bibitem[{Wilms {et~al.}(2000)Wilms, Allen, \& McCray}]{Wilms00a}
Wilms, J., Allen, A., \& McCray, R. 2000, The Astrophysical Journal, 542, 914,
  \dodoi{10.1086/317016}

\bibitem[{Yamasaki {et~al.}(2016)Yamasaki, Mizumoto, Ebisawa, \&
  Sameshima}]{Yamasaki16}
Yamasaki, H., Mizumoto, M., Ebisawa, K., \& Sameshima, H. 2016, Publications of
  the Astronomical Society of Japan, 68, 80, \dodoi{10.1093/pasj/psw070}

\bibitem[{Zhang {et~al.}(2018)Zhang, Wang, \& Zhu}]{Zhang18}
Zhang, J.-X., Wang, J.-X., \& Zhu, F.-F. 2018, The Astrophysical Journal, 863,
  71, \dodoi{10.3847/1538-4357/aacf92}

\bibitem[{Zoghbi {et~al.}(2012)Zoghbi, Fabian, Reynolds, \& Cackett}]{Zoghbi12}
Zoghbi, A., Fabian, A.~C., Reynolds, C.~S., \& Cackett, E.~M. 2012, Monthly
  Notices of the Royal Astronomical Society, 422, 129,
  \dodoi{10.1111/j.1365-2966.2012.20587.x}

\end{thebibliography}
\bibliographystyle{aasjournal}

\appendix 
\section{Additional plots}
\renewcommand\thefigure{\thesection.\arabic{figure}} 
\setcounter{figure}{0} 
Three additional plots are given in Appendix.
First, the MCMC fitting results for all the nine spectra ratios (Figure~\ref{fig:all_ratiofitting}).
Second, all the MCMC contour plots are shown in Figure~\ref{fig:all_contours}, where none of the parameters correlate with the velocity, which indicates that the velocity is independently determined. 
Finally, Figure~\ref{fig:all_specfit} shows all the spectral fitting results, which are successful including the complex UFO features.

\begin{figure*}
\centerline{\includegraphics[width=0.92\columnwidth]{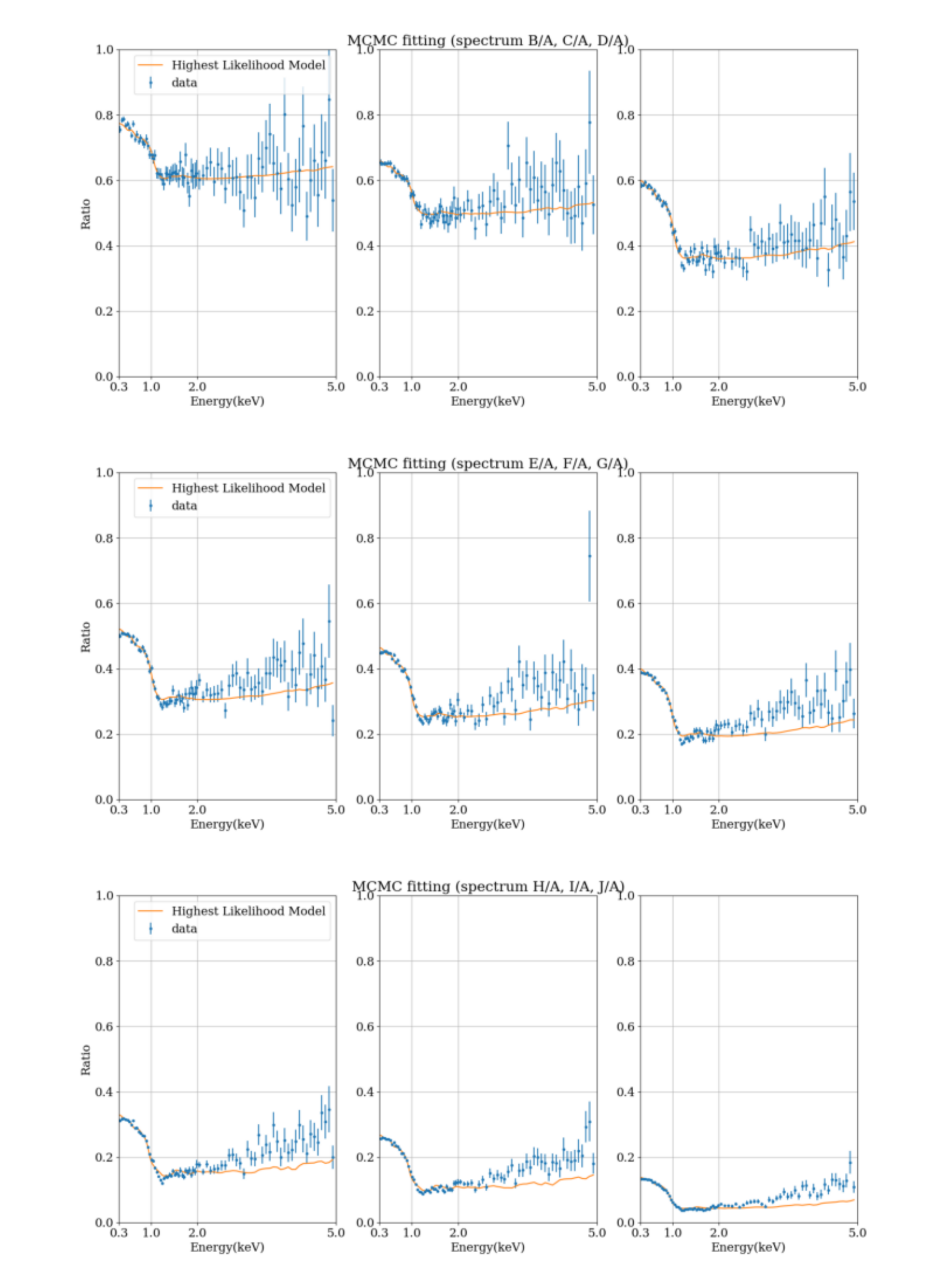}} 
	    \caption{MCMC fitting results of all the spectral ratios.}
	    \label{fig:all_ratiofitting}
\end{figure*}

\begin{figure*}
\centerline{\includegraphics[width=0.77\columnwidth]{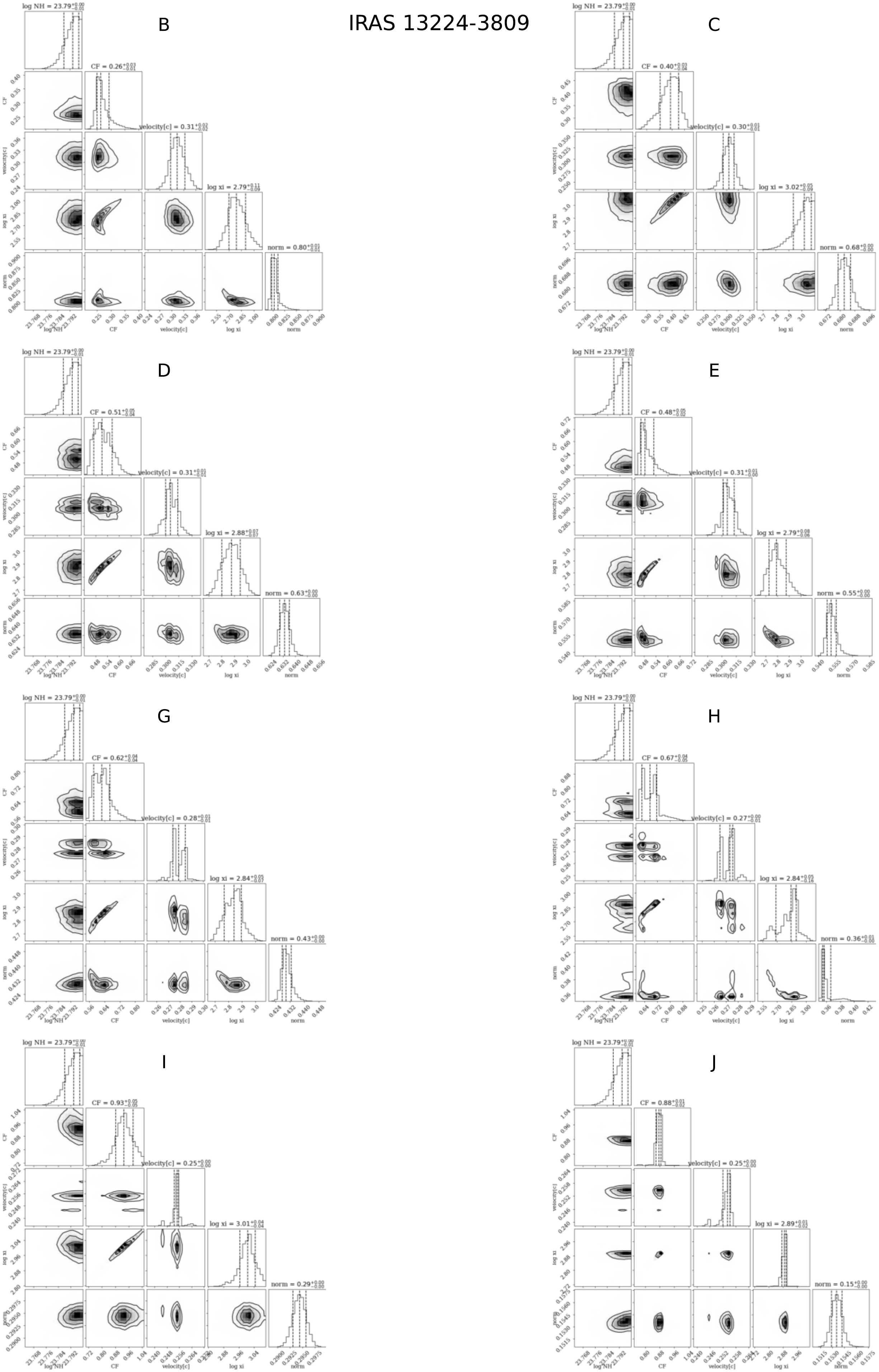}} 
	    \caption{Corner plots of the MCMC parameter estimation for all the spectral ratios except the spectral ratio F/A which is in the main text.
	 }
	    \label{fig:all_contours}
\end{figure*}

\begin{figure*} 
\centerline{\includegraphics[width=0.95\columnwidth]{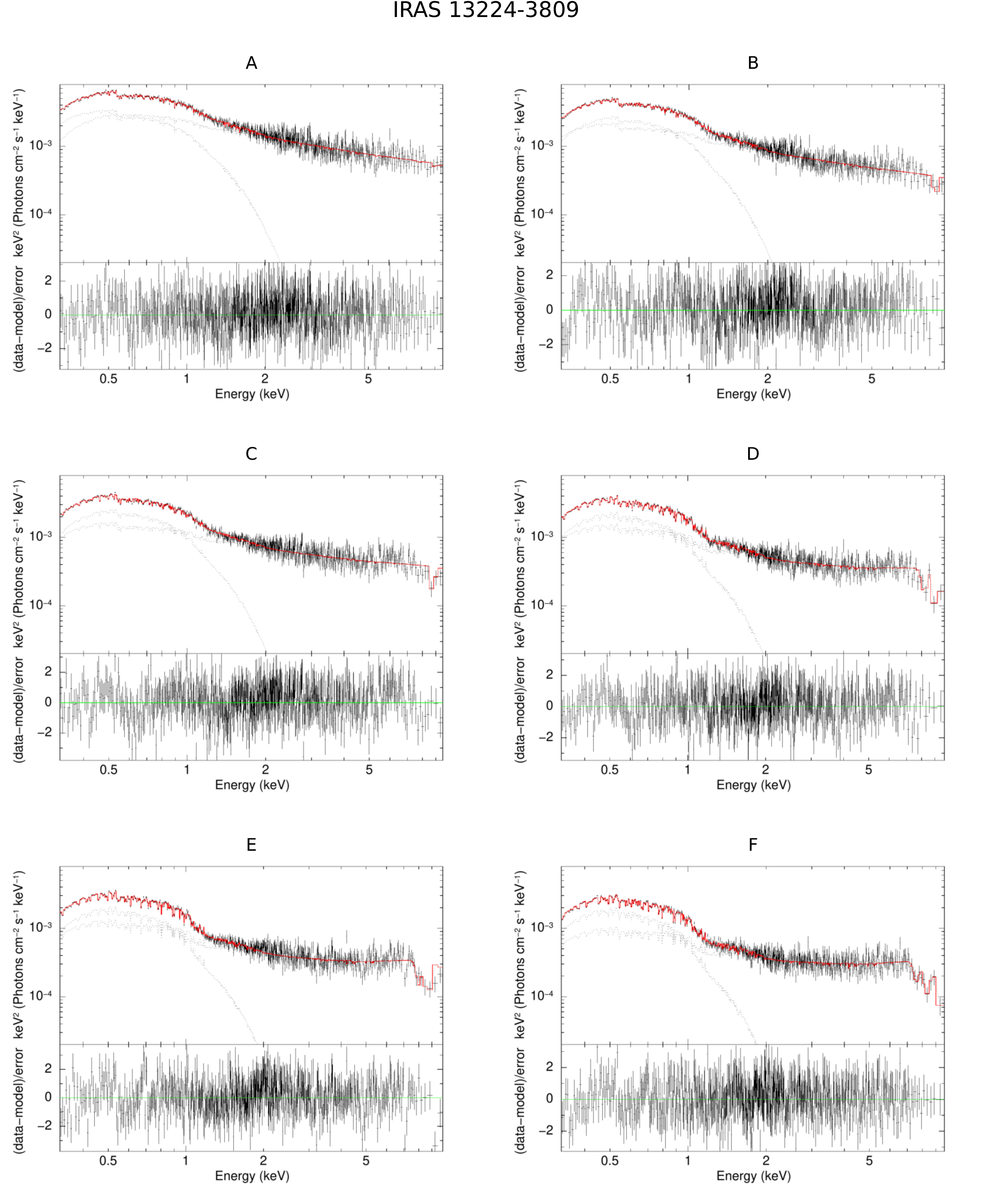}} 
	    \caption{Spectral fitting results of all the spectra except the spectrum F in the main text.}
	    \label{fig:all_specfit}
\end{figure*}

\begin{figure*} 
\centerline{\includegraphics[width=0.95\columnwidth]{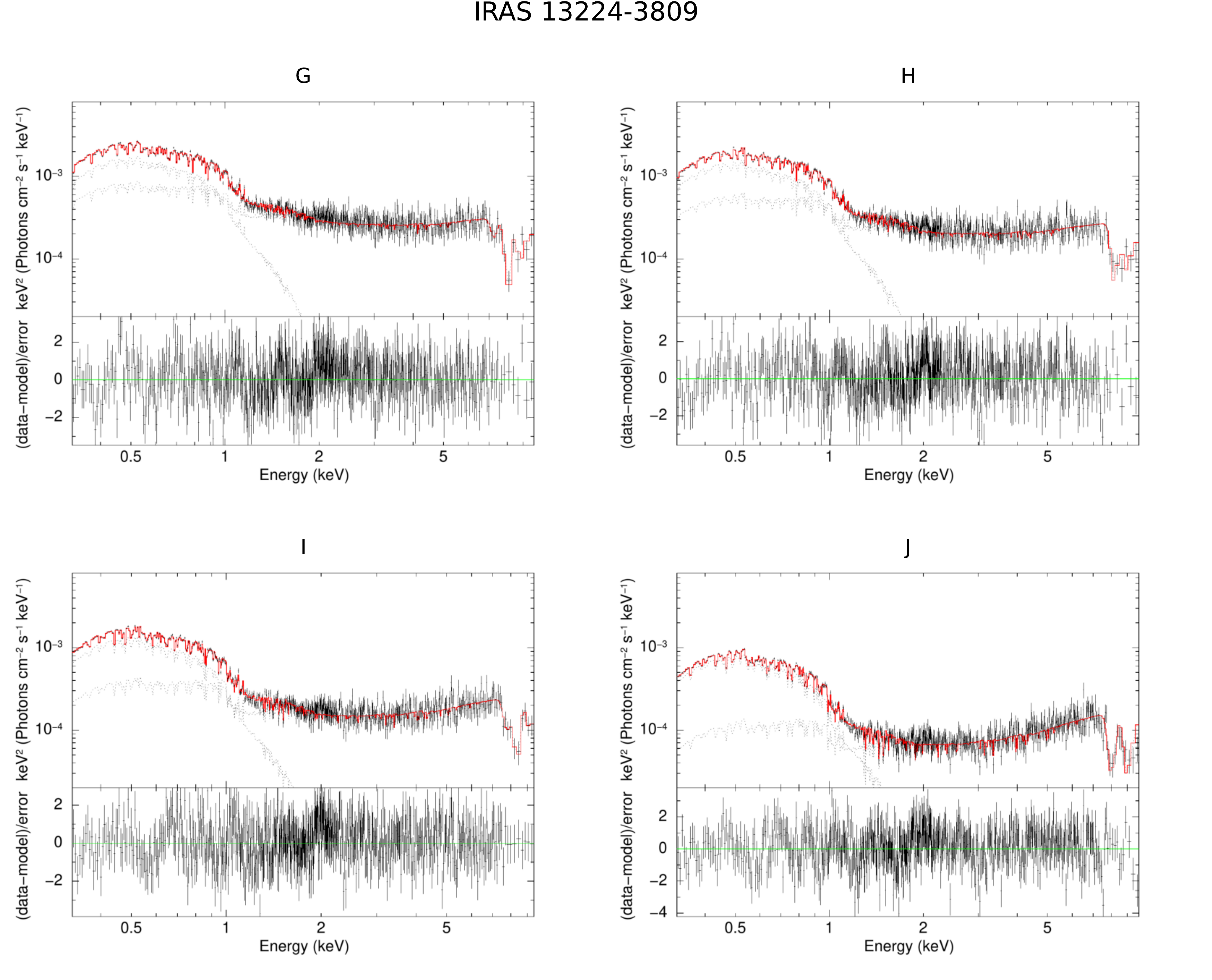}} 
	    \caption{(Cont.) Spectral fitting results of all the spectra except the spectrum F in the main text.}
\end{figure*}


\end{document}